\documentclass[%
 aip, amsmath,amssymb, reprint,
 jcp,twocolumn,10pt,floatfix,fleqn]{revtex4-2}

\usepackage{graphicx}
\usepackage{dcolumn}
\usepackage{amsmath}
\usepackage{amsfonts}
\usepackage{amssymb}
\usepackage{amsbsy}
\usepackage{bm}
\usepackage{mathrsfs}
\usepackage{upgreek}

\usepackage[utf8]{inputenc}
\usepackage[T1]{fontenc}
\usepackage{mathptmx}
\usepackage{etoolbox}

\usepackage{physics}

\usepackage{textcomp}
\usepackage{xspace}
\newcommand{\too}{2\texttimes1\texttimes1\xspace}
\newcommand{\oto}{1\texttimes2\texttimes1\xspace}
\newcommand{\TTo}{3\texttimes3\texttimes1\xspace}

\makeatletter
\def\@email#1#2{%
 \endgroup
 \patchcmd{\titleblock@produce}
  {\frontmatter@RRAPformat}
  {\frontmatter@RRAPformat{\produce@RRAP{*#1\href{mailto:#2}{#2}}}\frontmatter@RRAPformat}
  {}{}
}%
\makeatother
\begin{document}

%\preprint{AIP/EC-BU-2026}

\title[Scale-Bridging Reconstruction]
{Graph-Covariant Bottom-Up Scale-Bridging Reconstruction 
 of Ground and Excited States of Extended Systems}

\author{{Bartosz B{\l}asiak}}
\email{blasiak.bartosz@gmail.com}
\thanks{Corresponding author}
\affiliation{Institute of Physical and Theoretical Chemistry, 
             Goethe University Frankfurt, Max-von-Laue-Str. 7, 
             60438 Frankfurt, Germany}

\author{{Irene Burghardt}}
\affiliation{Institute of Physical and Theoretical Chemistry, 
             Goethe University Frankfurt, Max-von-Laue-Str. 7, 
             60438 Frankfurt, Germany}

\date{\today}

\begin{abstract}
Functional materials are multiscale disordered systems requiring
a `system-as-a-whole' level of description at both vibrational
and electronic state manifolds. 
We propose a hierarchical scale-bridging 
construction for building first-principles electronic structure models
of extended molecular aggregates with disorder, 
and its practical realization in exciton systems.
The general construction, referred here as to the \emph{encapsulation}, is based on a
recursive operator-valued hierarchy cast in a dual graph-theoretic (discrete) and
configuration-space (continuous) structure, in which variables associated with a
coarser level parameterize operators governing the degrees of freedom at
the next finer level in a graph-covariant, geometry-preserving fashion. 
We further exemplify the hierarchy
at three levels relevant for exciton systems: 
low-frequency collective motions of the aggregate, 
local vibrations of the monomers, and 
local electronic Hilbert spaces of 
coupled chromophores.
In an exemplary application of this scheme,
aggregate-level ground- and excited-state %vibronic 
operators %such as 
are constructed through a %n
many-body expansion (MBE), with %truncated at the two-body level, with
matrix elements evaluated in a basis comprising the aggregate ground state
and local Frenkel excitations. The required monomer and dimer building
blocks are obtained from DFT and TDDFT calculations, yielding
aggregate-level 1-particle transition density matrices as well as Hamiltonian,
dipole moment and other operators 
that can be evaluated at the level of the building blocks.
Starting from
structural models obtained from experiment or molecular dynamics, this approach
generates aggregate-level operators in a bottom-up fashion %vibronic 
%Hamiltonians 
without fitting parameters
to spectroscopic data or performing calculations on larger system cutouts. 
A numerical demonstration is given for an excitonic manifold in crystalline quaterthiophene.
The present realization employs MBE, DFT/TDDFT and local
Frenkel excitations, but the hierarchical construction is not restricted
to these choices and can accommodate other fragmentation methods, 
more general diabatic bases, as well as
multireference electronic-structure methods.
We anticipate that the framework proposed here 
can be used to consistently organize information
across multiple scales with broadly defined fragmentation
and coarse-graining
schemes. In particular, 
through its intrinsic configuration-dependence encoded in the encapsulations, it
%the dual hierarchy, 
will provide a bridge between static electronic
structure calculations for realistic, structurally disordered aggregates
and quantum or quantum-classical simulations of exciton--vibrational
dynamics. 
\end{abstract}

\maketitle
\tableofcontents

% =====================================================================================
%
\section{Introduction}
\label{sec:introduction}

Extended molecular systems such as molecular crystals, organic self-assemblies,
and (bio)polymeric architectures are important constituents of functional %molecular
materials
whose properties depend on the interplay between spatial or constitutional disorder
and their simultaneous organization 
across multiple spatial scales.\cite{Khasbaatar.etal.ChemRev_Assembly,
Popp.Brey.Binder.Burghardt.AnnuRevPhysChem.2021}
In fact, static and dynamic disorder can compete with 
long-range supramolecular organization
and strongly reshape charge and energy transport as well as optical 
response.\cite{Friedrich.Symalla.Meded.Neumann.Wenzel.JCTC.2014,
Illig.NatCommun.2016,Troisi.Orlandi.PhysRevLett.2006,
Wittmann.etal.JACS.2020}
Consequently, quantitative descriptions of both ground-state structure and dynamics
and excited-state electronic structure of molecular materials at distinct scales
are central to understanding and controlling
their function.\cite{Sharifzadeh.JPhysCondMat.2018,
Popp.Brey.Binder.Burghardt.AnnuRevPhysChem.2021}
Of particular interest are low-frequency lattice and quasi-acoustic motions
which
encode intermolecular forces and bulk thermoelastic response
\cite{Banks.etal.JMaterChemC_Thermoelasticity,Banks.Kleist.Ruggiero.NatRevChem_THz}
and 
were reported to modulate
electronic transport and exciton localization in optoelectronic materials.
\cite{Alvertis.etal.PhysRevB_ExcitonVibration,
Alvertis.etal.PRL_PhononLocalization}

At the electronic-structure level, direct 
%high-level 
supersystem calculations rapidly
become computationally prohibitive as the spatial extent and the number of relevant
electronic states increase.\cite{Grimme.Schreiner.AngewChemIntEd.2018}
A broad family of approximate methods has therefore been developed based on
fragmentation, embedding, and coarse-graining strategies, here referred generally
as to the electronic structure fragmentation methods (ESFs).
\cite{Gordon.etal.ChemRev_FragmentationReview,Oberhofer.Reuter.Blumberger.ChemRev.2017,
Loboda.etal.FrontiersChem.2023}
Prominent subsystem approaches include the many-body expansions (MBE)\cite{Hankins.Moskowitz.Stillinger.JCP.1970,Cisneros.etal.ChemRev.2016}
such as generalized many-body expansion
(GMBE),\cite{Richard.Herbert.JCP_GMBE,Richard.Herbert.JCTC_Overlapping}
the diabatic many-body expansion (DMBE),\cite{Paz.Glover.JCTC_DiabaticMBE}
density-based many-body expansion (D-MBE),\cite{Schuermann.etal.PCCP.2023}
and the fragment molecular orbital (FMO) method,
as well as
the generalized energy-based fragmentation (GEBF).
\cite{Li.Li.Jiang.JPCA_GEBF,Hua.etal.JPCA_GEBFVibrations,Fang.Jia.Li.JPCA_PBCGEBFVibrations,
Du.etal.JCTC_GEBFEmission,Li.etal.JPCA_GEBFExcited}
\cite{Fedorov.Kitaura.JCP_FockMBE}
Complementary strategies include divide-and-conquer (D{\&}C),
\cite{Yang.PRL.1991,Cabrera-Trujillo.Robles.PhysRevB.2001,
Kobayashi.Yoshikawa.Nakai.CPL.2010,Nishimura.Yoshikawa.Sakata.Nakai.JCTC.2026}
dynamical growth with molecular-orbital localization (DG-LMO),
\cite{Scemama.Renon.Rapacioli.JCTC.2014}
diabatic fragmentation schemes,
\cite{Cembran.Song.Mo.Gao.JCTC.2009,Oberhofer.Blumberger.PCCP.2012,
Lange.Voth.JCTC.2013,Valeev.etal.JACS}
systematic molecular fragmentation (SMF),\cite{Collins.PCCP.2012}
various embedding schemes,\cite{Wesolowski.Shedge.Zhou.ChemRev.2015,
Jones.Mosquera.Schatz.Ratner.JACS.2020,
Huevel.Reinholdt.Kongsted.JPCB.2023}
electron-density reconstruction (EDR),\cite{Exner.Mezey.JCC.2003}
and quantal-force-field approaches,\cite{Giese.etal.JCTC.2013,
Difley.etal.ACR.2010,Xie.Song.Truhlar.Gao.JCP.2008}
including hybrid quantum mechanics-molecular mechanics (QM/MM),\cite{Difley.etal.ACR.2010}
effective fragment potentials (EFP2),\cite{Gordon.Smith.Xu.Slipchenko.AnnuRevPhysChem.2013,
Nguyen.Pachter.Day.JCP.2014,
Day.Jensen.Gordon.Webb.Stevens.Krauss.Garmer.Basch.Cohen.JCP.1996}
X-Pol,\cite{Xie.Song.Truhlar.Gao.JCP.2008}
diatomics-in-molecules (DIM)\cite{Ellison.JACS.1963,
Olson.Garrison.JCP.1984,Belyaev.Tiukanov.Domcke.PhysRevA.2001,
Mills.etal.JPCB.2016} and others.
More recently, graph-theoretic fragmentation approaches
that naturally reflect long-range coarse-graining
have been developed to construct
multidimensional potential-energy surfaces\cite{Kumar.DeGregorio.Iyengar.JCTC_GraphPES}
and connect them via tensor-network representations with wavepackets 
and quantum nuclear dynamics.\cite{Kumar.etal.JCTC_TensorNetwork}

Beyond electronic structure fragmentation schemes, specialized computational %reduction 
strategies have also been developed for other demanding aspects 
of molecular simulation, particularly vibrational analysis 
of extended systems. It is well known that as the system size increases, constructing 
and diagonalizing the full atomistic Hessian becomes increasingly expensive
and at some point prohibitive due to insufficient memory. 
A further challenge arises from %the large variations 
%in 
the mechanical stiffness 
of extended molecular systems, 
where relatively soft collective deformations coexist with much stiffer local motions, 
giving rise to a broad range of vibrational frequencies.
These difficulties can be addressed 
in two complementary ways: by combining an electronic structure 
fragmentation (ESF) scheme with Hessian reconstruction, 
or by directly reducing the dimensionality and complexity 
of the vibrational problem. We collectively refer to this class of approaches as stiffness coarse-graining (SCG) 
methods.\cite{Tama.etal.Proteins_RTB,Ghysels.etal.JCTC_MBH,Houjou.JCTC_CoarseVibrations,
HuixRotllant.Ferre.JCTC_FragmentModes,Houjou.Seshimo.JMathChem_PBCPhonon,Wang.Seshimo.Houjou.JCTC_ISA,Soprani.etal.JCTC_MMD}

SCGs have quite a long history, initiated with 
calculations based on molecular mechanics force fields for proteins.
\cite{Mouawad.Perahia.Biopolymers.1993,Tirion.PRL.1996}
Direct normal mode analysis (NMA) of proteins by Brooks and Karplus\cite{Brooks.Karplus.PNAS.1983} was reported 
as early as in 1983,
and later reduced-dimensionality improvements were made 
by Mouawad and Perahia\cite{Mouawad.Perahia.Biopolymers.1993} in 1993
or Tama \emph{et~al.}\cite{Tama.etal.Proteins_RTB}.
The latter approach is especially relevant since it utilizes
a rotation and translation block (RTB) model of Durand \emph{et al.}\cite{Durand.Trinquier.Sanejouand.Biopolymers.1994}
casting motions of subunits within a rigid body picture.
At the same time, a distinct strategy was developed by
elastic network models \cite{Bahar.Atilgan.Erman.FoldingAndDesign.1997}
with explicit graph theory organization of coarse structure of extended systems
via Kirchhoff matrices by Bahar \emph{et~al}.\cite{Bahar.Atilgan.Erman.FoldingAndDesign.1997}
Newer methods allowed calculations at the electronic structure level.
Fang \textit{et al.} extended the GEBF method to molecular crystals 
to include periodic boundary conditions (PBCs) and reconstructed their vibrational 
spectra directly from nonperiodic calculations based on fragments.\cite{Fang.Jia.Li.JPCA_PBCGEBFVibrations} 
Building on the RTB idea, Houjou's group developed 
a coarse-grained description of intermolecular vibrations 
\cite{Houjou.JCTC_CoarseVibrations} 
and later extended this framework to periodic phonon calculations.\cite{Houjou.Seshimo.JMathChem_PBCPhonon} 
More recently, Wang, Seshimo, and Houjou introduced the independent stiffness approximation (ISA), 
in which an approximate dynamical matrix is assembled from small partial stiffness matrices 
within a coarse-grained molecular 
representation.\cite{Wang.Seshimo.Houjou.JCTC_ISA} 
Soprani \textit{et al.} 
formulated the minimal molecular displacement (MMD) approach, which employs a RTB-based construction
extended by a few selected low-frequency intramolecular modes
to substantially reduce the cost of crystalline phonon calculations.\cite{Soprani.etal.JCTC_MMD}

On the other hand, diabatic Hamiltonian reconstruction methods
(here collectively referred to as DHRs) provide an important route to
excited-state modeling in condensed phases.
\cite{Quarti.Fazzi.Tommasini.CPL_DensityMatrixExcitons,
Morrison.You.Herbert.JCTC_FrenkelDavydov,Fujita.Mochizuki.JPCA.2018,
Li.etal.JCTC_AIM,Singh.Roehr.JCTC.2024,
Adelsperger.Graaf.Roehr.JCTC.2026,
Pitesa.etal.JCTC_ECI,RodriguezMayorga.etal.JCTC_ExcitonMapping,
Santra.Bose.Ghosh.JCTC_SingletFission,
Blasiak.Maj.Cho.Gora.JCTC.2015,Blasiak.Bartkowiak.Gora.PCCP.2021,
Krueger.Scholes.Fleming.JPCB_TDC}
At an electronic %more general 
operator level, Fedorov and Kitaura formulated a
MBE of the Fock matrix,\cite{Fedorov.Kitaura.JCP_FockMBE}
while Paz and Glover introduced a MBE of a
charge-localized diabatic electronic Hamiltonian matrix elements.
\cite{Paz.Glover.JCTC_DiabaticMBE}
These methods can treat localized excited states.
However, to deal with delocalized excited states such as excitons,
an efficient route is to construct
a basis of local molecular excitations, charge-transfer states, and,
where necessary, multiexcitonic configurations.
An early fragment-based realization of this strategy was
the renormalized excitonic method (REM)
by Hajj, Malrieu and Guih\'ery\cite{Hajj.Malrieu.Guihery.PRB.2005},
later specialized to Frenkel Hamiltonian reconstruction by
Ma and Ma.\cite{Ma.Ma.JPCA.2013}
In REM, the MBE is used to approximate the Hamiltonian
based on subsystem excited states projected onto a common Frenkel basis,
which was later generalized to other electronic operators.\cite{Liu.Wang.Ma.CJCP.2021,Wang.Xie.Luo.Ma.JPCL.2022}
Another approach
was reported by Morrison, You, and
Herbert, who constructed an \textit{ab initio} Frenkel--Davydov
Hamiltonian in a basis of fragment excited states.
\cite{Morrison.You.Herbert.JCTC_FrenkelDavydov}
Within the FMO framework, Fujita and Mochizuki subsequently developed a
fragment-based treatment of nonlocal excitations in which local and
interfragment charge-transfer configurations enter an excited state
Hamiltonian constructed from fragment calculations.
\cite{Fujita.Mochizuki.JPCA.2018}
Similar strategy was used by Li \textit{et al.} to extend an \textit{ab initio}
exciton model beyond purely local Frenkel excitations by explicitly
including intermolecular charge-transfer states.
\cite{Li.etal.JCTC_AIM}
More recent developments have systematically enlarged the 
space of electronic configurations accessible to build effective excited state
Hamiltonians.
Pite\v{s}a \textit{et al.} introduced excitonic configuration interaction
(ECI), constructing aggregate Hamiltonians from monomer states and their
one-electron density matrices and allowing increasingly rich
multi-excitation manifolds.\cite{Pitesa.etal.JCTC_ECI}
For singlet-fission problems, Singh and R\"ohr extended fragment-orbital
model in the symbolic CI (SymCI) scheme developed in R\"ohr's group 
to spatially separated charge transfer and triplet pairs
\cite{Singh.Roehr.JCTC.2024}
while Adelsperger, de Graaf, and R\"ohr recently formulated a more general
framework encompassing one- and two-particle excitonic
manifolds.
\cite{Adelsperger.Graaf.Roehr.JCTC.2026}
Another similar strategy is to extract the reduced Hamiltonian by mapping
higher-level aggregate electronic states onto a physically motivated
diabatic basis rather than directly coming from the CI expansion. 
Rodriguez-Mayorga \textit{et al.}, for instance, developed
a multistate diabatization by projection that maps many-body \textit{ab initio}
states onto embedded Frenkel/charge-transfer Hamiltonians.
\cite{RodriguezMayorga.etal.JCTC_ExcitonMapping}
In particular, Santra \textit{et al.} demonstrated a
spin-resolved excitonic Hamiltonian parameterized from monomer and dimer
calculations and applied it to very large aggregates up to a hundred pentacene 
monomers and included PBCs
to handle crystals.\cite{Santra.Bose.Ghosh.JCTC_SingletFission}
Yet other, 
closely related %to these Hamiltonian reconstruction 
approaches are methods
rooted in perturbation theory
that obtain the intermolecular matrix elements themselves from transition
densities or reduced density matrices.
\cite{Krueger.Scholes.Fleming.JPCB_TDC,
Quarti.Fazzi.Tommasini.CPL_DensityMatrixExcitons,Fujimoto.JCP.2012,
Blasiak.Maj.Cho.Gora.JCTC.2015,Blasiak.Bartkowiak.Gora.PCCP.2021}
Such quantities can provide transferable microscopic ingredients for
constructing excitonic Hamiltonians in extended molecular assemblies
that are independent from the actual aggregate and close in spirit
to effective fragment potential philosophy. %such as EFP2, SolEFP and TrCAMM.

Despite this impressive variety of methods 
(ESFs, SCGs and DHRs that will be grouped here
simply as the `fragmentation methods' or FMs),
the corresponding algorithms are often highly procedural, largely independent and specialized
for particular observables or scales,
and, to our knowledge, there is no formal unified framework connecting 
them. 
This stays in contrast to condensed phase physics
where rigorous theories connecting strain and other constitutive mechanics
variables with electronic motion was established
via deformation potential theory as early as in 1950 and
received significant attention afterwards.\cite{Bardeen.Shockley.PhysRev.1950,
Manes.PhysRevB.2007,Pereira.CastroNeto.PRL.2009,Guinea.Katsnelson.Geim.NatPhys.2010,deJuan.Manes.Vozmediano.PhysRevB.2013}
Importantly, already at the mechanical level,
multi-scality necessarily induces a gauge structure of finer and coarser scale.\cite{Nguyen.Casale.LeMarrec.MatMechSol.2022}
To this end, from the purely FM perspective,
the coupling between long-range structural and compositional
disorder including collective motion and static material heterogeneity, on the one hand,
and excited-state manifolds and their properties in condensed phases, on the other,
has received comparatively little formal analysis. % from fragmentation perspective.
In particular, the geometric consequences of partitioning a molecular system
into spatial scales with fragments, including relationships between local
frames and global multi-scale assembly,
have not been systematically incorporated into a unified construction.

Here we attempt to propose such a rigorous geometric construction
that systematically bridges distinct spatial scales and 
electronic structure of ground and excited states,
linking them through a %unified 
hierarchy of graph-covariant representations
of scale transitions as a natural coarse-graining language. 
Within this common framework, collective coordinates,
transition densities,
%ground- and excited-state 
electronic Hamiltonians as well as ground- and excited-state properties
can be represented and reconstructed while remaining 
part of the same underlying principle across scales.
Of course, the purpose of the present theory is not to replace
existing ESF, SCG, or DHR methodologies. Rather, its modular structure is
designed to provide a common geometric scaffold through which such methods
can, in principle, be interfaced and their information transferred
consistently across scales. 
In view of this, the present contribution may be regarded as a geometric
`glue' for existing FMs, organizing their specialized information
while preserving and controlling the geometry of an extended or condensed phase system as a whole.

This paper is organized as follows. First, we present the general formalism based
on the dual continuous-discrete representation of scale transitions in
Sec.~\ref{sec:general}. Next, in Sec.~\ref{sec:theory}, we derive an exemplary practical
realization of this formalism to obtain 
collective modes and the excited-state Frenkel manifold of an
extended molecular system by selecting the well known 
MBE as a demonstrative FM.
This is then followed by a detailed discussion of the naturally
emerging gauge properties in Sec.~\ref{sec:gauge}. Periodic boundary conditions
are discussed in Sec.~\ref{sec:pbc}. We then validate the theory on a molecular
crystal in Sec.~\ref{sec:validation} and conclude in Sec.~\ref{sec:conclusions}.

% ======================================================================

\section{General Principle of Bridged Scales}
\label{sec:general}
% ======================================================================

Here we establish a general multiscale principle that governs the construction
of objects in this work. Let $\mathcal X^{(l)}$ be a configuration
space at scale level $l$ and $\mathcal H$ be the Hilbert space
of the full system at the finest electronic structure scale.
Then, the following hierarchy of scales is introduced as
\begin{equation} \label{e:basic}
 \mathcal X^{(l)}    \leftarrow 
 \mathcal X^{(l-1)}  \leftarrow \cdots \leftarrow
 \mathcal X^{(1)}    \leftarrow \mathcal H \;,
\end{equation}
where the arrow defines the mathematical structure induced by 
lifting the lower level configuration space to a higher level space.
The chain ends at the electronic structure level which is obtained
from lifting the highest structural scale to the Hilbert space.

\subsection{Scale-Bridging Constraints and Geometry}

\begin{figure}
%\begin{figure*}
   \centering
   \includegraphics[width=8.46cm]{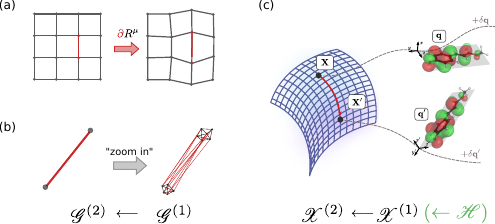}
   \caption{
  %\textbf{Geometry of the Bottom-Up Encapsulations for a Single Scale Transition.}
   \textbf{Geometry of a Single Scale Transition.}
   (a) Graph-covariant derivative describes the local deformation
   that results in global changes in the system's geometry;
   (b) Coarse graph edge lifted to fine graph space reveals
   microscopic details and inter-node constraints, e.g., a set of distances
   between atoms;
   (c) Example three-level hierarchy with the collective manifold as the base,
   local deformation as fibers lifted to Hilbert space in each
   $(X,q)$ configuration micro-state. The gauge freedom affecting 
   local vibrational eigenvectors as well as the electronic state phases
   and orientations is denoted as local coordinate systems.
   }
   \label{fig:geometry}
 \end{figure}

Sometimes it is cumbersome to work with continuous spaces $\mathcal X^{(l)}$,
especially when defining or reconstructing coarse-grained variables
and collective motions, both being more naturally described by discrete variables.
This is because coarse-grained systems often obey a discrete set of constraints
that are tightly associated with their %fragmented in nature 
topology, such as stiffness constraints
in fragmented systems where the constraints are energetically allowed ranges 
of relative distances and orientations
of the fragments or subdomains.
In turn, continuous representation of these same
quantities can be preferable, e.g., in classical or quantum dynamics schemes.
Therefore, here we associate each configuration space level
with its corresponding discrete graph representation, $\mathcal G^{(l)}$,
which obeys an analogous hierarchy as in Eq.~\eqref{e:basic}, i.e.,
\begin{equation} \label{e:basic-g}
 \mathcal G^{(l)}    \leftarrow 
 \mathcal G^{(l-1)}  \leftarrow \cdots \leftarrow
 \mathcal G^{(1)}    \;.
\end{equation}
Under a given set of discrete constraints encoded in each graph
level, the coarse scale manifold 
can be defined through pulling back its metric tensor from
the finer scale, i.e.,
%composed 
%of the graph-covariant derivatives of the fine level coordinate,
%
\begin{equation} \label{e:basic-g-metric}
 g_{ij}^{(l)} = \left[ 
                     \nabla^{\mathcal G^{(l)}} {\bf X}^{(l-1)} 
                \right]_i
                \cdot
                \left[ 
                     \nabla^{\mathcal G^{(l)}} {\bf X}^{(l-1)} 
                \right]_j
 \;,
\end{equation}
where $\nabla^{\mathcal G^{(l)}}$ denotes
a continuous derivative on $\mathcal X^{(l)}$ induced by lifting the
incidence structure and constraints of $\mathcal G^{(l)}$ into the
tangent space %bundle 
of the corresponding configuration manifold.
We shall
refer to it as the \emph{graph-covariant derivative}
which describes a change of variable while respecting global constraints
imposed by the slice of the graph hierarchy from Eq.~\eqref{e:basic-g} up to $\mathcal G^{(l)}$,
as illustrated in Fig.~\ref{fig:geometry}(a).
Related approaches in discrete differential geometry have 
also employed graph edges and vertices to define connections 
and covariant geometric operators, establishing links between such discrete representations 
and the corresponding operators on smooth manifolds.\cite{Dimakis.MuellerHoissen.JMP.1994,Singer.Wu.CPAM.2012,Bandeira.Singer.Spielman.SIMAX.2013,
Burago.etal.MathZ.2022,Jiang.GaugeGraphs.2023}
The approach developed here, however, differs from these formulations 
in that the graph is not used to represent the configuration 
space itself, but rather to describe structural constraints which, 
when lifted to the underlying continuous molecular configuration space, induce the corresponding derivative.

The charts on $\mathcal X^{(l)}$ can be defined by eigenvalue decomposition
of ${\bf g}^{(l)}$ that encode natural orthogonal directions
and their distance measures, along with possibly redundant or ill-defined sets of directions.
Such directions might be due to an over-complete description
of the finer scale (which is to be discarded) 
and/or manifest constraint frustrations in the system
that are physical and associated with mechanical memory or hysteresis effects.
The graph representation of Eq.~\eqref{e:basic-g} 
is therefore crucial to define the allowed geometry 
of lower-level configuration states that is limited by constraints
defining the material under study. The prerequisite is the knowledge
of the finest structural resolution of the system described by ${\bf X}^{(1)}$.

The above organization of geometric information is analogous in spirit to
previous reduced mechanical descriptions where local (internal) degrees 
of freedom form fibers over a reduced configuration manifold and changes of local frames 
necessarily induce geometric connections.\cite{Littlejohn.Reinsch.RevModPhys.1997,Nguyen.Casale.LeMarrec.MatMechSol.2022}
It also relates to approaches which
explicitly employ the collective scale metric tensor dependence at the quantum level.\cite{Tokatly.PRB.2007}
However, here we generalize this already broad concept to work on a dual continuous/discrete object %(encapsulation)
that represents scale transitions as a `zooming-in' process of coarse, discrete information
into a continuous representation at a finer scale, thus fitting naturally into atomistic simulations. 
Fig.~\ref{fig:geometry}(b) schematically shows such a process of a `lift' in discrete graph space between two scales,
$\mathcal G^{(2)} \leftarrow \mathcal G^{(1)}$, 
where $\mathcal G^{(1)}$ can be simultaneously embedded in its associated configuration space $\mathcal X^{(1)}$
such as atomic coordinate space as exemplified in Fig.~\ref{fig:geometry}(c).
Moreover, we allow several such scale transitions
to be connected with each other into a global %encapsulation 
continuous-discrete fibration structure, ultimately ending
with the Hilbert space.

%\end{figure*}

In this work, we explicitly apply the above construction
to define collective mode configuration space directly from Cartesian
configuration space as ${\bf X}^{(1)}$ via the associated graph-covariant derivative
under discrete microscopic stiffness constraints.
Nevertheless, the dual discrete-continuous construction is anticipated
to be useful for general types of coarse-grainings.

\subsection{Scale-Bridging Operators}

The scale chain in Eq.~\eqref{e:basic} induces hierarchical
dependence of any scale-specific local observable $\hat{O}^{(l)}$, i.e.,
\begin{equation} \label{e:basic-o}
 \hat{O}^{(l)} = \hat{O}^{(l)} \left[ \mathcal X^{(l+1)} \right] \;,
\end{equation}
where the exact dependence is up to this point arbitrary and can 
be chosen from the pool of FMs discussed in Sec.~\ref{sec:introduction}. 
Note that these operators can describe quite general observable features, 
which could be constitutive, geometric, vibrational, electronic or vibronic in nature,
depending on the problem in focus.
The definition of the dual graph chain 
requires that an assembly %there must exist an assembly
of this operator at the associated discrete hierarchy level can be formulated, i.e., 
\begin{equation} \label{e:basic-og}
\hat{O}^{(l)}(\mathcal G) \equiv \mathsf{As} \{ \hat{O}^{(l)} | \mathcal G^{(l+1)} \leftarrow \mathcal G^{(l)} \}
\;,
\end{equation}
where $\mathsf{As}$ denotes a graph-assembly operation of local operators
$\hat{O}^{(l)}$ defined on $\mathcal G^{(l)}$ lifted from the local base $\mathcal G^{(l+1)}$.
Note that the graph assembly is therefore done explicitly on the latter (coarse) graph
and its form is strictly governed by the fragmentation method choice.

The graph-covariant derivative introduced in Eq.~\eqref{e:basic-g-metric} 
extends naturally to local operators defined at each scale.
Application of the chain rule on the graph assembly gives the 
operator derivatives along local coordinates,
\begin{multline} \label{e:basic-og-der}
   \nabla_{\mu}^{\mathcal G^{(k)}}
   \hat O^{(l)}(\mathcal G)
   =
   \left(
   \nabla_{\mu}^{\mathcal G^{(k)}}
   \mathsf{As}
   \right)
   \left\{
   \hat O^{(l)}
   \,\middle|\,
   \mathcal G^{(l+1)}
   \leftarrow
   \mathcal G^{(l)}
   \right\}
   \\
   +
   \mathsf{As}
   \left\{
   \sum_a
   \left[ {\bf J}^{(l+1\leftarrow k)} \right]_{\mu a}
   \nabla_{a}^{\mathcal G^{(l+1)}}
   \hat O^{(l)}
   \,\middle|\,
   \mathcal G^{(l+1)}
   \leftarrow
   \mathcal G^{(l)}
   \right\}
   \;,
\end{multline}
which is particularly useful for dynamical systems.
In the above equation, we defined the scale-bridging Jacobian
as given for the nearby scales as
\begin{equation}
{\bf J}^{(k-1\leftarrow k)} 
\equiv
\nabla^{\mathcal G^{(k)}}
{\bf X}^{(k-1)} \;,
\end{equation}
and a subsequent chain rule over distant scales as
\begin{equation} \label{e:jac-index}
 {\bf J}^{(m\leftarrow k)}
 =
 {\bf J}^{(m\leftarrow m+1)} \cdot
 {\bf J}^{(m+1\leftarrow m+2)}
 \cdots
 {\bf J}^{(k-1\leftarrow k)}
\end{equation}
for $k>m$.
Note that for coordinate-independent assembly
the first term in Eq.~\eqref{e:basic-og-der} vanishes.

\subsection{Scale-Bridging Dynamics}

Let $\mathcal L^{(l)}$ be a dynamical generator of coordinates ${\bf X}^{(l)}\in \mathcal X^{(l)}$.
Then, via Eq.~\eqref{e:basic}, their dynamics is dependent upon all coarser scales,
%
%\begin{equation} \label{e:basic-dx}
$ \partial_t X^{(l)} = \mathcal L^{(l)}
 \left[
    {\bf X}^{(l)}; {\bf X}^{(l+1)}, \ldots
 \right] .
$
%\end{equation}
%%
At the highest (quantum) level the generator is the aggregate's Hamiltonian,
\begin{equation} \label{e:basic-dh}
 {\rm i}\hbar \partial_t \ket{\Psi} = \mathscr{H} 
 \left[
  {\bf X}^{(1)}, {\bf X}^{(2)}, \ldots
 \right]
 \ket{\Psi} \;.
\end{equation}
In fact, Eq.~\eqref{e:basic} implies that with a given choice
of local frames as well as electronic structure fragmentation method, 
the scale transition mappings
define the lower-level differential manifolds,
which makes it possible to quantize these degrees of freedom
and re-cast the microscopic Hamiltonian of the aggregate
(or any other operator as long as it can be computed via chosen ESF
for a chosen set of vertices and edges) in a multiscale fashion.
In this work we show 
an example of such a process in Sec.~\ref{subsubsec:collective_mode_definition}
when defining quantum collective displacement observables.
On the other hand, in mixed quantum-classical setups,
certain configuration variables are more convenient to be kept
in the classical sector. 
For example, one can straightforwardly define the scale-bridged positions
${\bf Q} \equiv \{ {\bf Q}^{(l)}, \ldots \}$ for ${\bf Q}^{(l)} \in \mathcal X^{(l)}$ 
and the associated conjugate momenta ${\bf P} \equiv \{ {\bf P}^{(l)}, \ldots \}$
and propagate them %the system %including the classical variables ${\bf Z} = \{ {\bf Q}, {\bf P}_{{\bf Q}}\}$
via existing dynamical methods. We shall discuss such a classical setup 
from the scale-bridging perspective in Sec.~\ref{subsec:vibronic_coordinates}.

\subsection{Encapsulations}

For brevity, we shall refer to the dual hierarchical graph-covariant construction
$(\mathcal X | \mathcal G ; \mathcal H)$ as to \emph{encapsulations} (EC),
to emphasize 
that the locally defined and mutually separable
building blocks are preserved, together with the mappings describing how their representations are connected across different scales. While these building blocks remain local and independent at their respective scale levels, the constraints imposed on them, as well as their induced structural responses, are carried through
the hierarchy by the graph-covariant scale transition maps, thus influencing the overall organization of the system. 
In this construction, the graph spaces are used to describe the discrete constraints and connectivity between the local objects, whereas the corresponding continuous configuration spaces provide their actual molecular realization. The interplay between these two structures then allows for successive lifts connecting 
the local structural degrees of freedom with the collective organization at larger scales and, ultimately, with the associated electronic Hilbert space.

In the following sections, we develop a first realization
of the above general construction, targeting exciton
systems. The resulting scale-bridging vibronic coupling models can be combined with
a fully non-adiabatic quantum dynamical level of description.

% ======================================================================
\section{Three-Level Graph-Covariant Encapsulation of Exciton Systems}
\label{sec:theory}
% ======================================================================

\begin{figure}
   \centering
   \includegraphics[width=0.45\textwidth]{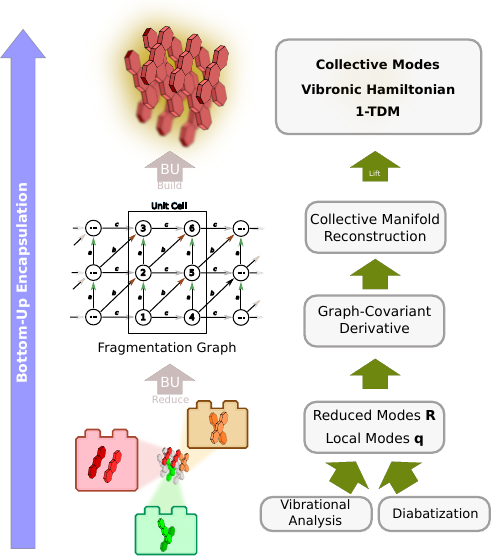}
   \caption{
   \textbf{Three-Level Bottom-Up Encapsulation of Extended Aggregates.}
   The system is reduced to selected dimer building blocks that are assembled
   together according to a fragmentation graph to build the aggregate model.
   The procedure distinguishes local and collective vibrational modes, 
   built systematically from supramolecular calculations on dimers
   which, after graph-covariant derivative calculation and collective
   manifold reconstruction yield aggregate-level observables. 
   }
   \label{fig:ec_bu}
 \end{figure}

In this section, we formulate the bottom-up 
transition-density many-body expansion (TrMBE)
of the aggregate-level Hilbert space that is based on 
a three-scale instance of Eq.~\eqref{e:basic}, 
\begin{equation} \label{e:basic-3}
 \mathcal X_{\rm Collective}    \leftarrow 
 \prod_I \mathcal X^{(I)}_{\rm Local}  \leftarrow 
 \bigotimes_I \mathcal H^{(I)}_{\rm el}
\end{equation}
where the base of the hierarchy $\mathcal X_{\rm Collective}$ 
is the inter-fragment
collective motions of monomers leading to large-scale
deformations within the material.
The method is schematically summarized in Fig.~\ref{fig:ec_bu}.
Particular atomistic detail is resolved at $X^{(I)}_{\rm Local}$
for the $I$th monomer in the aggregate. At the end, 
electronic structure information is constructed
by the lift of the composite Hilbert space generated
by a tensor product of local monomer Hilbert spaces, $\mathcal H^{(I)}_{\rm el}$.
We choose the well-known MBE\cite{Hankins.Moskowitz.Stillinger.JCP.1970} as the electronic fragmentation method,
truncated at the two-body level (here referred to as MBE-2),
and
applied directly to the electronic density and Hamiltonian 
operators in the fragment-diabatic state basis.\cite{Ma.Ma.JPCA.2013}
Note that 
the general GMBE methodology by Richard and Herbert\cite{Richard.Herbert.JCP_GMBE,Richard.Herbert.JCTC_Overlapping} can be used
for covalently bound species, which
reduces to a simple MBE for non-covalently bound species discussed here.
To this end, 
the required electronic
structure information is obtained from monomer and dimer building blocks, 
chosen here for simplicity 
at the density functional theory\cite{Hohenberg.Kohn.PR.1964,Kohn.Sham.PR.1965} (DFT; ground-state level) and 
the linear-response
time-dependent density functional theory\cite{Casida.1995,Stratmann.Scuseria.Frisch.JCP.1998} (LR-TDDFT; excited states level),
assuming localized Frenkel excitations.
Here, we point out that the dimensionality of the collective
and local configuration space can be reduced via known effective 
mode constructions,\cite{Tamura.Bittner.Burghardt.JCP_Hierarchy,Popp.etal.JCP_EffectiveModes}
thus making the scheme applicable at a fully non-adiabatic quantum dynamics level
of Eq.~\eqref{e:basic-dh}
when extended to a vibronic Hamiltonian via derivatives
with respect to the collective and local modes (see Ref.~\citenum{Blasiak.etal.JCTC_PBI}).
At this level, existing high-dimensional
wavepacket propagation methods such as the
Multi-Layer-(Gaussian)-Multi-Configurational Time-Dependent Hartree (ML-(G)MCTDH) methods
\cite{Wang.Thoss.JCP.2003,Manthe.JCP.2008,Vendrell.Meyer.JCP.2011,Romer.Ruckenbauer.Burghardt.JCP.2013,DiMaiolo.Worth.Burghardt.JCP.2021}
can directly provide
equations of motion for the dynamics.

In particular, the proposed scheme allows to approximately reconstruct
the aggregate's transition density, transition dipole moments 
and vibronic Hamiltonian which, once combined with quantum dynamics
or quantum-classical dynamics schemes, yields generic tools to model
exciton dynamics in extended aggregates with complex molecular topology
and disorder. %and transition dipole moment operators,
To our knowledge, 
in contrast to %reduced molecular-coordinate 
%and 
coarse-grained constructions of collective crystal 
vibrations\cite{Houjou.JCTC_CoarseVibrations,Houjou.Seshimo.JMathChem_PBCPhonon,Wang.Seshimo.Houjou.JCTC_ISA,Soprani.etal.JCTC_MMD} 
and fragment-based constructions of effective excitonic 
Hamiltonians\cite{Morrison.You.Herbert.JCTC_FrenkelDavydov,Pitesa.etal.JCTC_ECI,
RodriguezMayorga.etal.JCTC_ExcitonMapping,Santra.Bose.Ghosh.JCTC_SingletFission}, 
ECs place both descriptions within a common graph-covariant hierarchy.
Importantly, local structural responses are lifted 
to a nonredundant aggregate collective manifold 
while coordinate-dependent ground- and excited-state operators,
such as reduced transition density matrix and Hamiltonian, 
are reconstructed across 
\emph{the same} unified hierarchy of structural and electronic scales.

Before we proceed, we establish indexing notation
used throughout the remainder of this work. Fragments are labeled as $I$, $J$, $K$,
graph edges as $e$ or $f$,
atom IDs as Greek letters as $\alpha$ or $\beta$, local and collective vibrational
modes are labeled via $r$, $s$ and $i$, $j$, $k$, respectively,
with the exception of collective modes in isolated dimers
which are labeled as $\mu$ or $\nu$ instead.  
Atomic orbital (AO) indices are labeld as $\xi$ or $\eta$.
Diabatic and adiabatic electronic states are labeled by $n$ or $m$
and $a$ or $b$, respectively. Cartesian components corresponding
to $x$, $y$ and $z$
are denoted by $u$, $w$ and $z$.

% ----------------------------------------------------------------------
\subsection{Fragmentation Graph Space}
\label{subsec:fragmentation_graph_space}
% ----------------------------------------------------------------------

\subsubsection{Two-Body Truncation Ansatz}
Consider an aggregate composed of $N$ chromophoric units. Its fragmentation is
represented by a coarse-scale graph referred here as the \emph{fragmentation graph},
\begin{equation}
  \mathcal G = (\mathcal V,\mathcal E)
  \qquad
  \text{ for }
  \mathcal V = \{1,\ldots,N\},
\end{equation}
where each vertex $I\in\mathcal V$ denotes one monomer and each edge
$e=(I,J)\in\mathcal E$ denotes a dimer building block included in the
%two-body 
MBE-2 expansion of the local electronic operator\cite{Fedorov.Kitaura.JCP_FockMBE}
\begin{equation} \label{e:mbe2}
 \hat{O}(\mathcal G) = \sum_I (1-\mathsf D_I) \hat{O}_I + \sum_e \hat{O}_e
 \;,
\end{equation}
where $\mathsf D_I$ is the degree of the $I$th node, measuring the amount of 
interactions with other nodes, whereas $\hat{O}_I$ and $\hat{O}_e$ are the superimposed
operators of the $I$th isolated fragment and $e$th isolated dimer, respectively.
The factor $(1-\mathsf D_I)$ ensures that no double counting of the monomer contributions is done.
Note that the above expansion notation emphasizing graph structure %employing graph degree matrix
is equivalent to standard expansions reported in the past for non-convalently bound species. %TODO: cite \cite{Fedorov.Kitaura.JCP_FockMBE}
In Eq.~\eqref{e:mbe2}, the locality (here fragment-diabaticity) of the electronic operator
is crucial since otherwise the reconstruction would not be accurate due to delocalization.\cite{Ma.Ma.JPCA.2013}

% ......................................................................
\subsubsection{Representable Monomers and Dimers}
\label{subsubsec:average_dimer_concept}
% ......................................................................

For extended aggregates, the number of geometrically distinct dimers can be very
large. 
To reduce the computational cost of the MBE-2 setup, % --> MBE-2 coined in the Introduction!
the graph is %also 
supplemented by vertex/edge-type maps
\begin{subequations}\label{e:sets_t}
\begin{align}
  \tau_1:\mathcal V &\rightarrow \mathcal T_1 \;, \\
  \tau_2:\mathcal E &\rightarrow \mathcal T_2 \;, 
\end{align}
\end{subequations}
where $\mathcal T_1$ and $\mathcal T_2$ is the set of structurally distinct monomer and dimer classes
in the aggregate, %Exemplary fragmentation graphs are shown in Fig.~\ref{f:graphs}(a)
as well as vertex/edge-symmetry maps,
\begin{subequations}\label{e:sets_s}
\begin{align}
  \sigma_1:\mathcal V &\rightarrow \mathcal R_1 \;, \\
  \sigma_2:\mathcal E &\rightarrow \mathcal R_2 \;, 
\end{align}
\end{subequations}
containing symmetry operations on a given vertex/edge type.
Thus, the Cartesian geometry of a monomer and dimer 
element in the graph is reduced to
\begin{subequations}
\begin{align}
{\bf Q}_v &= \mathcal S[v\leftarrow \tau_1(v)] \left(\sigma_1(v)  {\bf Q}_{\tau_1(v)} \right) \;,\\
{\bf Q}_e &= \mathcal S[e\leftarrow \tau_2(e)] \left(\sigma_2(e)  {\bf Q}_{\tau_2(e)} \right) \;,
\end{align}
\end{subequations}
where $\mathcal S$ is the Cartesian coordinate superimposition map described later.
Such proxy representatives defined for each vertex and 
edge type
$t_p\in\mathcal T_p$ will be here referred as to the ``representative dimers'' (RDs),\cite{Ploetz.Niehaus.Kuehn.JCP.2014}
and ``representative monomers'' (RMs).
A similar strategy was already introduced previously, e.g., in the EFP2-type superimposition machinery
of benchmark fragments (monomers)\cite{Gordon.etal.ChemRev_FragmentationReview,Blasiak.Ritchie.Webb.Cho.PCCP.2016} 
and in crystals for dimer species,
\cite{Ploetz.Niehaus.Kuehn.JCP.2014,Berkelbach.Hybertsen.Reichman.JCP.2014,Rivera.Dommett.Sidat.Rahim.Crespo-Otero.JCC.2020}
and is here formalized to our graph construction framework.
For instance,
for all $(I,J)\in\mathcal E$ 
we will denote the type of each dimer as 
$t_2=\tau(I,J)$,
with 
the corresponding RD geometry denoted by
${\bf Q}_{t_2}$. Electronic structure calculations are performed for
these representative dimers 
only and the resulting matrix elements are assigned to
all graph edges of the same type according to the formulae derived later in Sec.~\ref{subsec:electronic_hamiltonian}.
Thus, the additional controlled approximation
amounts to replacing 
the edge-resolved 
summations
by the type-resolved summations,
$\sum_e \approx \sum_t \sum_{e\in t}$
and $\hat{O}_{e} \approx \hat{O}_{\tau(e)}$.
Here, $\hat{O}_{\tau(e)}$ denotes the two-body operator defined for the
RD of type $\tau(e)$.
The exact same design as described above for dimers is also applied for monomers.
In fact, as will be evident from the forthcoming sections,
this controlled approximation is necessary to make the calculations
of the aggregate's vibronic Hamiltonian
feasible.

 \begin{figure*}
   \centering
   \includegraphics[width=0.75\textwidth]{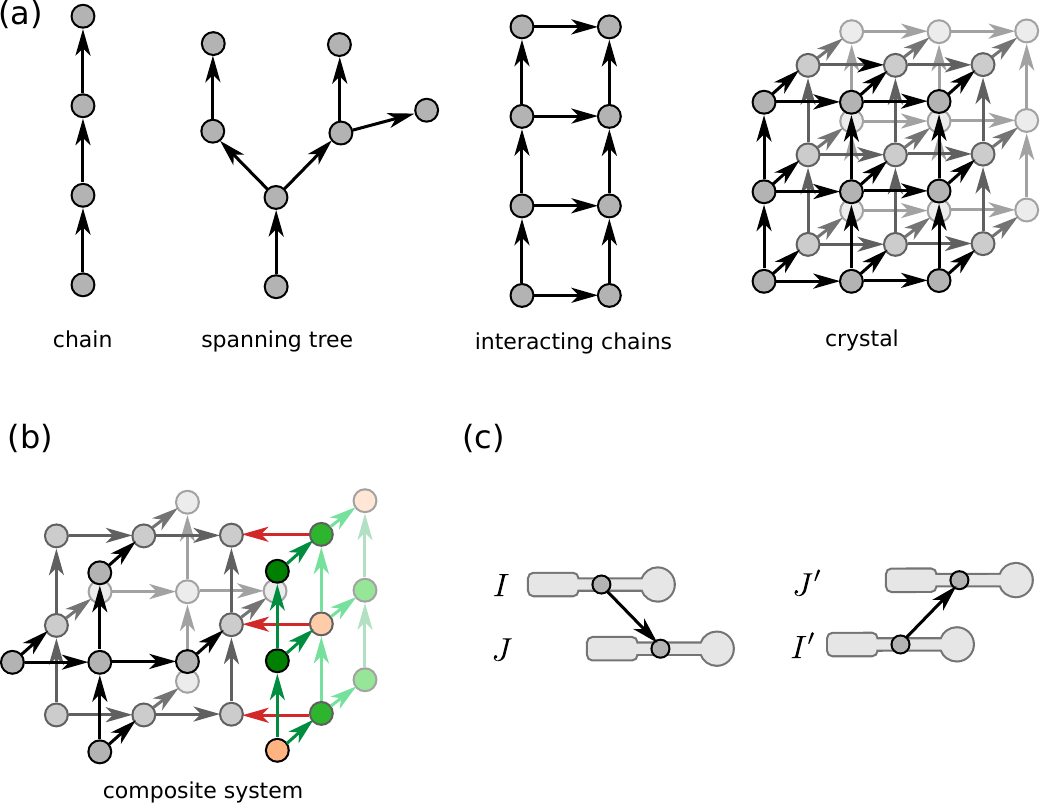}
   \caption{\textbf{Fragmentation Graphs Corresponding to the Collective Mode Manifold
   as the Base of the Three-Scale Hierarchy.}
   (a) Exemplary topologies and their directed graphs. Each vertex and edge can have arbitrary types
   depending on compositional (vertex-related) and structural (edge-related) disorders.
   (b) Example of a multi-domain host-guest system with two interacting different graph classes 
   and few monomer types denoting matrix and dopants.
   (c) Edge directionality distinguishes monomers even when they are locally identical.
   The directionality must be preserved among equivalent dimer classes such as 
   in the two shown dimers.
   }
   \label{fig:aggregate_topologies}
 \end{figure*}

% ......................................................................
\subsubsection{Fragmentation Graphs}
\label{subsubsec:fragmentation_graphs}
% ......................................................................

Fig.~\ref{fig:aggregate_topologies}(a) shows basic aggregate topologies
that can be represented in the form of fragmentation graphs.
Apart from simpler topologies of chains and spanning trees, 
networks are particularly important %that contain cycles 
since they describe many important classes
of realistic chemical systems from crystals and nanosheets to 
%anisotropic 
fibrous self-assemblies
and disordered aggregates in solution.
Higher order organizations of few interacting topologies 
such as hybrid structures with compositional disorder 
in Fig.~\ref{fig:aggregate_topologies}(b)
are particlularly straightforward to represent using fragmentation
graphs.
Note that the graph directionality is physical, i.e., it reflects
the microscopic relationships between two interacting dimers, exemplified
in Fig.~\ref{fig:aggregate_topologies}(c).
However, the graph orientation itself is used as a bookkeeping convention for
ordered dimer construction and lattice displacement assignment. We emphasize that the physical
MBE-2 Hamiltonian is independent of this convention, provided that reverse edges
are treated consistently (see also Sec.~\ref{subsec:graph_directionality}). 
It is also important to mention that not all monomers in the graph
must be electronically excited, i.e., the electronic quantum numbers $n$ could be constrained
to $0$ in a sub-set of monomers.
This situation corresponds to a nearest-neightbour model within MBE-2
for the molecular environment around the electronically active quantum system.

% ----------------------------------------------------------------------
\subsection{Multiscale Coordinates}
\label{subsec:vibronic_coordinates}
% ----------------------------------------------------------------------

The aggregate Hamiltonian is parameterized by collective coordinates of the
aggregate denoted by $\{\Xi_j\}$, 
local intramolecular coordinates of the monomers denoted by $\{{\bf q}_{I}\}$, 
and a fragment-diabatic
electronic basis, 
with $\ket{G}$ being the common ground state
and $\{\ket{\bf I}, \ket{\bf IJ}, \ldots\}$
being the diabatic states associated with excitations
of monomers, dimers, and so on. Although the theory
applied here is applicable to any type of fragment-diabatic basis,
for clarity of presentation we limit the discussion
to the Frenkel basis, $\{\ket{I_n}\}$,
and mark that extensions of the basis set by adding 
other states (e.g.,
charge-transfer or correlated triplet states) follow straightforwardly based
on the forthcoming sections.

The three levels of description, i.e, collective, local atomic and electronic 
are necessarily hierarchical,
i.e., 
\begin{equation} \label{e:hierarchy}
  \bm{\Xi} \longleftarrow
  {\bf q}[\bm{\Xi}] \longleftarrow
  \ket{G[{\bf q}[\bm{\Xi}]]} \,,\, 
  \ket{{\bf I}[{\bf q}[\bm{\Xi}]]} \,,\ldots
\end{equation}
with the collective mode manifold
as the base of the multi-scale hierarchy. Note the parametric dependences
of the smaller scales on their parent scales. We choose
the well-known Kabsch rigid-body superimposition map,\cite{Kabsch.ActaCrystA.1976} $\mathcal S[\mathrm{target} \leftarrow \mathrm{ref}]({\bf r}_\alpha) = {\bf r}_\alpha'$
that transforms the Cartesian position vector of atom $\alpha$, ${\bf r}_\alpha$,
by superimposing the reference structure onto a target structure.
The Cartesian coordinates of the aggregate will be denoted as ${\bf Q} = \{ \bm{\Xi}, {\bf q}\}$.
We shall also assume only one type of RM and explicitly highlight RD types
in the equations below merely for notational clarity, thus we denote $t_2\equiv t$
from now on
and omit summations over $t_1$. However, it is understood
that the theory below is defined for arbitrary number of monomer types
and the notation extension to monomer heterogeneity is straighforward. %but rather formidable.
In the next sections, we define the construction of each particular scale in detail.

% ......................................................................
\subsubsection{Collective Modes}
\label{subsubsec:collective_mode_definition}
% ......................................................................

Collective modes describe low-frequency motions of the aggregate, including
intermolecular and quasi-acoustic deformations. Due to the aggregate's prohibitive
size it is not possible to calculate them directly using standard electronic
structure methods.
Instead, we make use of the harmonic approximation
of the MBE-2 expansion of the ground state
potential,
\begin{multline}
 V_G({\bf Q};\mathcal G) = \sum_I (1-\mathsf D_I) v_I({\bf q}_I) 
              + \sum_e v_e({\bf R}_e; {\bf q}_I, {\bf q}_J)
 \\
 \approx \frac{1}{2}\sum_t\sum_{\mu\in t} m_{t_\mu}
       \omega_{t_\mu}^2 \sum_{e\ni t} \left( R_{e_\mu}    - R_{t_\mu,{\rm eq}}\right)^2 
  = V_G({\bf R};\mathcal G)
 \;,
\end{multline}
where ${\bf R}_e = {\mathcal S}[e\leftarrow t]({\bf R}_t)$ and
${\bf R}_t$ is a set of \emph{reduced vibrational dimer modes} (RVDMs)
which constitute collective vibrations within the isolated $t$-th RD corresponding
to the $e$th edge after superimposition ${\mathcal S}$, with
the reference RVDM transformation
\begin{equation} \label{e:d}
 {\bf d}_{\alpha t} = \frac{\partial {\bf r}_\alpha}{\partial {\bf R}_t} 
\;.
\end{equation}
RVDMs can be, for example, rigid body modes, or include
additionally intramolecular low-frequency deformations
such as monomer bending, twisting and so on.
Such modes can be found by reduced dimensional vibrational analysis\cite{Nagy.Vikar.Lendvay.PCCP.2018,Tobias.Arendas.Csaszar.JCTC.2022,ZepedaHerrera.etal.PCCP_Pentacene}
(see also Appendix~\ref{a:flex}).
The harmonic approximation in Eq.~\eqref{e:d} 
is expected to be sufficient, since 
in an aggregate the stiffness restricts molecules
from deforming far away from their local equilibria.

Note that the set of RVDMs is necessarily redundant and non-orthogonal.
Thus, to get the non-redundant and orthogonal collective modes
we seek for a transformation 
\begin{equation}\label{e:c_def}
 {\bf C}_{ej}(\mathcal G) = \frac{\partial {\bf R}_e}{\partial \Xi_j} 
 \;,
\end{equation}
where $\Xi_j$ span an unknown lower dimensional manifold.
We define it by eigendecomposition of the metric tensor
in RVDM sub-space,
\begin{multline}
g_{e_\mu f_\nu}({\bf R};\mathcal G) \equiv 
 \sum_\alpha
 \left[ {\bf d}_{\alpha e}({\bf R};\mathcal G) \right]_\mu\cdot 
 \left[ {\bf d}_{\alpha f}({\bf R};\mathcal G) \right]_\nu
 \\
 = \sum_j U_{e_\mu j} ({\bf R})
          U_{f_\nu j} ({\bf R})
          \lambda_j   ({\bf R})\;,
\end{multline}
which gives orthogonal collective directions weighted
by the distance measures $\lambda_j$. 
The null-space that corresponds
to the redundant degrees of freedom
(and, numerically, all the eigenvectors having measures
less than a chosen threshold $\lambda^*$)
should be discarded leading to the chart ${\bf X}$ with
its transformation
\begin{equation}
 F_{e_\mu j}({\bf R};\mathcal G) 
   \equiv \frac{\partial R_{e_\mu}}{\partial X_j({\bf R})} = 
             \frac{U_{e_\mu j}({\bf R})}{\sqrt{\lambda_j}({\bf R})} \;.
\end{equation}
Here, 
${\bf d}_{\alpha e}(\mathcal G)\equiv {\nabla}_e^{\mathcal G} {\bf r}_\alpha$
is the aggregate-level counterpart of the isolated dimer transformation ${\bf d}_{\alpha t}$
from Eq.~\eqref{e:d}, and
${\nabla}_e^{\mathcal G}$ is the fragmentation-graph-covariant derivative
that reflects the fact that displacing a selected RVDM in a given 
dimer (edge) necessarily changes positions and orientations
of all the surrounding connected monomers due to the material stiffness
propagated by the graph's edges as illustrated in Fig.~\ref{fig:geometry}(a).
To define an effect of such a derivative on ${\bf r}_\alpha$, the graph space
must be lifted to Cartesian coordinate space.
The simplest realization of such a lift is
performing a symmetric distortion along a chosen RVDM
while keeping all other RVDMs and the associated interatomic distances
constant. This leads to
\begin{equation}\label{e:cov_der}
 \left[ {\nabla}_e^{\mathcal G} {\bf r}_\alpha \right]_\mu ({\bf R})
 \equiv \left[ {\bf B}^+(\mathcal G; {\bf R}) \cdot {\bf b}_{e_\mu } (\mathcal G; {\bf R}) \right]_\alpha \;,
\end{equation}
where `$^+$' denotes the pseudo-inverse operation,
${\bf B}$ is the lift of the graph incidence to Cartesian coordinate space
defined for each microscopic structural edge constraint $l\ni e=(I,J)$ between two atoms $\alpha\in I$ and $\beta\in J$
in the following way:
\begin{subequations}\label{e:B}
\begin{align}
 \Delta_{\alpha\beta} {\bf B}_{l \alpha} &= {\bf r}_\alpha - {\bf r}_\beta \;,\\
 \Delta_{\alpha\beta} {\bf B}_{l \beta } &= {\bf r}_\beta  - {\bf r}_\alpha\;,\\
                      {\bf B}_{k \alpha} &= 0 \text{ for $k\neq l$} \;,
\end{align}
\end{subequations}
$\Delta_{\alpha\beta}$ is the interatomic distance, and
\begin{equation}\label{e:b}
 \left[ {\bf b}_{e_\mu} \right]_k =
 \begin{cases}
 \Delta_{\alpha\beta}^{-1}
 \left( {\bf r}_\alpha - {\bf r}_\beta \right) \cdot
 \left( {\bf d}^{(I)}_{\alpha e_\mu}
      - {\bf d}^{(J)}_{\beta e_\mu} \right),
 & k = l, \\[0.8em]
 0,
 & k \neq l ,
 \end{cases}
\end{equation}
with ${\bf d}^{(I)}_{\alpha e_\mu} = \mathcal S [e\leftarrow t]({\bf d}^{(I)}_{\alpha t_\mu}) $.
Once the covariant derivative and thus the collective chart ${\bf X}$
are defined, the Hessian follows immediately
from
\begin{multline}\label{e:hess_j}
%\begin{equation}\label{e:hess_j}
  H_{ij}({\bf R};\mathcal G) 
         \equiv \nabla_i^{\mathcal G} 
                \nabla_j^{\mathcal G}
                V_G({\bf R}) 
  \\
  \approx
  \sum_t \sum_{\mu\in t}
  m_{t_\mu} \omega_{t_\mu}^2 \sum_{e\ni t} 
  F_{e_\mu i} ({\bf R})
  F_{e_\mu j} ({\bf R})
  \;,
\end{multline}
where
\begin{equation}
\nabla_j^{\mathcal G} ({\bf R})  
 \equiv \sum_e \sum_{\mu\in e} 
  F_{e_\mu j} ({\bf R})
 \partial_{e_\mu} \;.
\end{equation}
The approximation made in Eq.~\eqref{e:hess_j} 
neglects geometrical terms 
related to frame mismatch away from the reference at ${\bf R}$ 
and also global intrinsic curvature of the collective manifold itself.
The diagonalization of the mass-weighted Hessian 
\begin{equation}
 {\bf H}^{\rm MW}(\mathcal G) = {\bf M}^{-\frac{1}{2}} \cdot {\bf H} \cdot {\bf M}^{-\frac{1}{2}}
\end{equation}
yields
the final collective mass-weighted normal modes and the associated vibrational frequencies,
\begin{equation} \label{e:coll_freq}
 H^{\rm MW}_{ij}(\mathcal G)
  = \sum_a \Omega_a^2 U_{ia}^{\rm MW} U_{ja}^{\rm MW} 
\end{equation}
with
the mass matrix in orthogonal ${\bf X}$ chart given by
\begin{equation}
 M_{ij}(\mathcal G) = \sum_e \sum_{\mu\in e} m_{e_\mu} F_{e_\mu i} F_{e_\mu j} \;.
\end{equation}
This also defines the transformation from Eq.~\eqref{e:c_def}
to be
\begin{equation} \label{eq:C_final}
 {\bf C}_{e_\mu j}(\mathcal G) = \sum_k F_{e_\mu k} 
 \left[  {\bf M}^{-\frac{1}{2}} \cdot {\bf U}^{\rm MW} \right]_{kj} \sqrt{\nu_j}
 \;,
\end{equation}
where the reduced masses of the collective modes are given by
\begin{equation}
 \nu_j(\mathcal G)  = \left[        \sum_{\alpha     } \lvert {\bf u}_{\alpha j}       \rvert^2 \right]^{-1}
 \;,
\end{equation}
and the transformation
\begin{multline}
\label{e:u_final}
 {\bf u}_{\alpha j}^{\rm MW}(\mathcal G) \equiv \frac{\partial {\bf r}_\alpha}{\partial \Xi_j^{\rm MW}}
 \\
 = \sum_{e} 
  \mathcal S[e\leftarrow t] \left( {\bf d}_{\alpha t} \right)
 \left[
  {\bf F} \cdot {\bf M}^{-\frac{1}{2}} \cdot {\bf U}^{\rm MW}
 \right]_{ej}
\end{multline}
gives the delocalized movement of all the fragments within the aggregate
along a specified collective mode,
rigorously encoding the transition from the large scale motion (${\bf U}^{\rm MW}$)
that branches the movement through anisotropic graph-resolved stiffness 
(${\bf F}$) towards particular atomic
deformations from superimposed ${\bf d}_{\alpha t}$.

Finite collective displacements then
require a path- or order-prescription, e.g. applying local monomer or edge
displacements in a specified sequence and propagating them through the graph.
The transformation in Eq.~\eqref{eq:C_final} is of critical importance 
since it maps isolated dimer quantities via their RVDMs to connected macroscopic 
system motions $\bm{\Xi}$. Therefore, quantities such as IR or Raman intensities,
transition dipole moment derivatives and vibronic couplings along the collective modes
can all be staightforwardly mapped to the underlying quantities of isolated RDs.

It remains to comment on the modelling assumption made
in the graph theoretic ansatz, i.e., the graph is rigid and 
its lift to Cartesian space is 
free from geometric frustrations, i.e., 
$[\nabla_j^{\mathcal G} , \nabla_k^{\mathcal G}] = 0$.
In such an ideal case, the deformation
of particular monomers will be consistent for each overlap
in the dimer connections creating zero residual strain
(e.g., for a connection 1--2--3,
the dimer 1--2 displacement and dimer 2--3 displacement will generate
identical movement of the monomer 2).
However, this is most certainly not possible in real condensed phase
congested system where mechanical frustrations are natural and lead
to path-dependence of the deformation 
propagation and accumulation of residual strain.\cite{Li.Semnani.CMAME.2025,Xu.Guo.Guo.Zhang.Wang.Polymers.2023,Hulm.Ochsenfeld.JCTC.2023}
This means that Eq.~\eqref{e:u_final} should be understood
under a mechanical gauge choice
that selects the pathway of the deformation inside the aggregate,
by applying local edge/vertex displacements in a specified sequence
and propagating them through the graph along a chosen path.
This gauge must be encoded already at the level of the graph-covariant
derivative from Eq.~\eqref{e:cov_der} and used consistently throughout the calculation if one needs to
visualize the distorted aggregate.
In this work we neglect the frustration problem since it exceeds the scope
of the present work. Hence, it is emphasized that Eqs.~\eqref{e:B}
and Eq.~\eqref{e:b} define the graph-covariant derivative 
under frustrationless conditions.

% ......................................................................
\subsubsection{Local Modes}
\label{subsubsec:local_mode_definition}
% ......................................................................

Local modes $\{{\bf q}_I\}$ describe intramolecular vibrations of individual monomers
that are in relative positions and orientations given by the collective
coordinates $\bm{\Xi}$ from the previous section as follows from
Eq.~\eqref{e:hierarchy}.
Thus, the natural definition of the microscopic displacement of the system
is
\begin{equation}
{\bf Q}({\bf q}[\bm{\Xi}];\mathcal G) 
 = %{\bf Q}_0 + {\bf u}({\mathcal G}) \delta \bm{\Xi} 
  {\bf Q}_0(\bm{\Xi};\mathcal G) 
  %\\
  + \sum_t \sum_{I\in t} 
 {\bf L}_{t(\bm{\Xi})}^{(I)}
 \cdot \delta {\bf q}^{(I)} 
 \;,
\end{equation}
where
\begin{equation} \label{e:multiscale_displacement}
 {\bf Q}_0(\bm{\Xi};\mathcal G) \equiv {\bf Q}_0({\bf 0}) 
 + {\bf u}({\mathcal G})\cdot \delta \bm{\Xi}
\end{equation}
is the geometry with only collective displacements
activated relative to the reference geometry of the model
${\bf Q}_0({\bf 0})$ (or simply ${\bf Q}_0$ for brevity),
${\bf u}$ is the Cartesian-to-collective mode
transformation ${\bf u}_{\alpha j} \equiv {\bf u}_{\alpha j}^{\rm MW} \sqrt{\nu_j}$,
${\bf L}_{t(\bm{\Xi})}^{(I)}=\mathcal S[I({\bf Q}_0(\bm{\Xi}))\leftarrow I({\bf Q}_0({\bf 0}))]
\left({\bf L}_{t(0)}^{(I)}\right)$,
and ${\bf L}_{t(0)}^{(I)}$
are the local normal modes superimposed onto ${\bf Q}_0$.
The reference geometry is defined as the starting geometry
(possibly from experiment, molecular dynamics simulation
or other source)
that was optionally subjected to energy optimization with respect to the collective 
displacements following $-\nabla^{\mathcal G} V_G$ to reach 
a collective stationary state. 

It is also instructive to analyze the dynamical properties of
the above hierarchical ansatz. The microscopic velocity follows as
\begin{equation} \label{e:multiscale_velocity}
 \frac{{\rm d}{\bf Q}}{{\rm d}t} \cong
 {\bf u}(\mathcal G) \cdot 
  \frac{{\rm d}\bm{\Xi}}{{\rm d}t}
 %\\
 + \sum_t \sum_{I\in t}
 {\bf L}_{t(\bm{\Xi})}^{(I)}
 \cdot
 \frac{ {\rm d}_I(\mathcal G) {\bf q}_I } {{\rm d}t}
  \;,
\end{equation}
where 
\begin{equation} \label{e:multiscale_time_derivative}
 \frac{ {\rm d}_I(\mathcal G) {\bf q}_I } {{\rm d}t}  \equiv \frac{{\rm d}{\bf q}_I}{{\rm d}t}
 + \sum_j {\bf A}_{jI} (\mathcal G) \cdot {\bf q}_I 
   \frac{{\rm d}\Xi_j}{{\rm d}t}
\end{equation}
is the covariant time derivative
with the connection approximately defined by $  {\bf L}_{t(\bm{\Xi})}^{(I)} \cdot{\bf A}_{jI}= \nabla^{\mathcal G}_j {\bf L}_{t(\bm{\Xi})}^{(I)}$,
and we assumed that $\nabla^{\mathcal G}_j {\bf u} \approx {\bf 0}$
(see Appendix~\ref{sec:gauge_potential_collective} for more details).
This defines the canonical momenta as
\begin{subequations} \label{e:multiscale_canonical_momenta}
\begin{align}
 {\bf p}_I(\mathcal G) &= {\bf g}_I \cdot \frac{ {\rm d}_I(\mathcal G) {\bf q}_I } {{\rm d}t} \;,\\
 {\bf P}  (\mathcal G) &= {\bf g}_{\bm{\Xi}} \cdot \frac{ {\rm d} \bm{\Xi} } {{\rm d}t}  + {\bf Y}(\mathcal G) 
 \;,
\end{align} 
\end{subequations}
where the local and collective metric tensors are
cast for unscaled normal mode convention as
\begin{subequations} \label{e:multiscale_metric}
\begin{align}
 {\bf g}_I &= \{{\bf L}^{(I)}_{t(\bm{\Xi})}\}^{\mathsf T} \cdot {\bf m}_I \cdot {\bf L}^{(I)}_{t(\bm{\Xi})} \;,\\
 {\bf g}_{\bm{\Xi}} &= {\bf u}^{\mathsf T} \cdot {\bf m} \cdot {\bf u} \;,
\end{align} 
\end{subequations}
${\bf m}$ is the standard diagonal atomic mass matrix,
and the kinematic connection is
\begin{equation} \label{e:multiscale_kinematic_connection}
 {Y}_j(\mathcal G) \equiv \sum_I {\bf A}_{jI} (\mathcal G) :  {\bf p}_I(\mathcal G) \otimes {\bf q}_I  \;.
\end{equation}
Additionally, one finds that collective velocity is
\begin{equation} \label{e:collective_velocity}
 \frac{ {\rm d} \bm{\Xi} } {{\rm d}t} = {\bf g}_{\bm{\Xi}}^{-1} \cdot \left( {\bf P}  (\mathcal G) - {\bf Y}  (\mathcal G) \right) \;.
\end{equation}
The above result is important as it reveals
dynamical coupling between the collective and local vibrational
modes that was induced by the multiscale design despite the coordinates
$\bm{\Xi}$ and ${\bf q}$ being orthogonal to each other. This is relevant
for the definition of the kinetic energy operator
discussed later in Sec.~\ref{subsubsec:kinetic_energy}.

% ......................................................................
\subsubsection{Electronic States}
\label{subsubsec:electronic_state_definition}
% ......................................................................

The diabatic electronic basis consists of the common aggregate ground state
$\ket{G}$ and the set of local Frenkel excitations per monomer,
\begin{equation}
  \ket{I_n({\bf q}[\bm{\Xi}];\mathcal G)}
  =
  \ket{n_I({\bf q}_I[\bm{\Xi}];\mathcal G)}
  \bigotimes_{J\neq I}
  \ket{0_J({\bf q}_J[\bm{\Xi}];\mathcal G)}
  \,.
\end{equation}
We use the above formal definition to derive matrix elements
of the aggregate's Hamiltonian and their geometry dependence
by fixing the internal coordinates at ${\bf q}_0$ of the basis and
the collective state at $\bm{\Xi}_0$.
However, note that despite the fact that the Frenkel basis states
are constant in their local frame defined by each monomer's position
in the aggregate,
the displacement along the collective modes leads to translation
and rotation of the electronic basis, leading to the frame drift effect,
\begin{multline}\label{e:state_displacement}
 \ket{I_n({\bf q}[\bm{\Xi}_0+\delta\bm{\Xi}];\mathcal G)}
 = 
  \mathcal S[I(\delta)\leftarrow I(0)]
  \left( \ket{n_I}\right) 
  \\
  \bigotimes_{J\neq I}
  \mathcal S[J(\delta)\leftarrow J(0)]
  \left( \ket{0_J}\right) 
  \;,
\end{multline}
where $\mathcal S$ denotes here a superimposition lift to the Hilbert space.

In this work, we rather use the one-particle reduced transition density matrices (1-TDMs)
to describe the electronic state properties such as the exciton density
or oscillator strengths, as it is more straightforward computationally 
within the LR-TDDFT framework. %TODO: citation to Neugebauer papers
At the operator level,
the electronic density operator can be written as
\begin{equation}
 \hat{\rho}_{\rm el}(\mathcal G) = \sum_I (1-\mathsf D_I) \hat{\rho}_{{\rm el},I} 
 + \sum_e \hat{\rho}_{{\rm el},e} 
 \;.
\end{equation}
Then, the aggregate-level Frenkel 1-TDM at ${\bf Q}_0$ is
\begin{multline} \label{e:tdm}
 \hat{\gamma}_{GI_n}(\mathcal G)  \equiv
 {\rm Tr}^{(1)} \left\{
 \hat{\rho}_{\rm{el} }(\mathcal G)\ket{I_n}\bra{G}
 \right\} %({\bf r}_1,{\bf r}_1')
 \\
 = (1-\mathsf D_I) 
 \mathcal S[I\leftarrow I_0]\left(
 \hat{\gamma}_{0n}^{(0)}
 %({\bf r}_1,{\bf r}_1') 
 \right) 
  \\
  + \sum_t \sum_{e\ni I}^{\in t}
 \mathcal S[e\leftarrow t]\left(
 \hat{\gamma}_{00,\pi_I[0n_I]}^{(t)}
 %({\bf r}_1,{\bf r}_1')  
 \right)
 \;,
\end{multline}
where $\hat{\gamma}_{0n}^{(0)}$ and $\hat{\gamma}_{00,\pi_I[0n_I]}^{(t)}$
are the 1-TDMs of the RM and $t$-th RD, respectively,
\begin{subequations}
\begin{align}
 \hat{\gamma}_{0n}^{(0)} &\equiv
 {\rm Tr}^{(1)} \left\{
 \hat{\rho}_{{\rm el},0}\ket{n_I}\bra{0}
 \right\} \;,\label{e:1tdm_m}\\
 \hat{\gamma}_{00,n_I0}^{(t)} &\equiv
  {\rm Tr}^{(1)} \left\{
 \hat{\rho}_{{\rm el},e}\ket{n_I0_J}\bra{0_I0_J}
 \right\} \;, e=(I,J) \label{e:1tdm_e1}\;,\\
 \hat{\gamma}_{00,0n_I}^{(t)} &\equiv
  {\rm Tr}^{(1)} \left\{
 \hat{\rho}_{{\rm el},e}\ket{0_Jn_I}\bra{0_J0_I}
 \right\} \;, e=(J,I) \label{e:1tdm_e2}\;,
\end{align}
\end{subequations}
${\rm Tr}^{(1)}$ traces out all spin and spatial electronic degrees of freedom 
except for spatial degrees of freedom of electron `1',
and the superimposition lifts to the 1-TDM space originate 
from the lifts to the state space in Eq.~\ref{e:state_displacement}.
Appendix~\ref{sec:rigid_molecule_algorithm} 
gives explicit formulae for the superimposition lifts of 1-TDM in the AO representation
that were used in the calculations.
The operator $\pi_I$ selects the appropriate permutation of the 
monomer indices in the dimer depending whether the excited particle 
lies in the first or second monomer (note that the edges are directional as
explained in Sec.~\ref{subsubsec:average_dimer_concept}).
It is worth clarifying that while the 1-TDMs of the RM from Eq.~\eqref{e:1tdm_m} are
directly accessible from the electronic structure calculations
as simply the adiabatic 1-TDMs of the monomer,
that is not the case for the 1-TDMs of the RDs from Eqs.~\eqref{e:1tdm_e1} and \eqref{e:1tdm_e2}
which %need to be computed
need to be computed via diabatization by projection
%need to be computed via projection-based quasi-diabatization
with respect to the relevant adiabatic dimer 1-TDMs,
\begin{equation} \label{e:tdm_ad}
\begin{pmatrix}
 \hat{\gamma}_{00,01_I}^{(t)} \\ \vdots \\
 \hat{\gamma}_{00,0n_I}^{(t)} \\ \vdots \\
 \hat{\gamma}_{00,1_I0}^{(t)} \\ \vdots \\
 \hat{\gamma}_{00,n_I0}^{(t)} \\ \vdots \\
\end{pmatrix}
 =
 {\bf U}_t \cdot
\begin{pmatrix}
 \hat{\gamma}_{00,1}^{(t)} \\ \vdots \\
 \hat{\gamma}_{00,a}^{(t)} \\ \vdots
\end{pmatrix}
\end{equation}
where ${\bf U}_t$ is the diabatic-to-adiabatic transformation matrix
of the isolated dimer corresponding to edge of type $t$
and $\hat{\gamma}_{00,a}^{(t)}$ is the adiabatic dimer 1-TDM
associated with the $a$th transition.
This approach follows the overlap maximization strategy
originally proposed by Cimiraglia~\emph{et al.}\cite{Cimiraglia.etal.JPB.1985}
and subsequently extended to the TDDFT framework
using one-particle transition densities by Neugebauer~\emph{et al.}\cite{Neugebauer.JCP.2004}.
The transformation matrix ${\bf U}_t$ is determined
by maximizing the overlap between the adiabatic dimer
states and the reference Frenkel states using L\"owdin symmetric orthogonalization 
by Yaghoubi Jouybari~\emph{et al.}\cite{YaghoubiJouybari.etal.JCTC.2020}.

An important consequence of Eq.~\ref{e:tdm} is the definition
of the aggregate-level transition dipole moments associated with the Frenkel states,
$\pmb{\upmu}_{0\square} = -{\rm Tr}\{\hat{\gamma}_{0\square}\,\hat{\bf r}\}$
which leads to an analogous formula:
\begin{multline} \label{e:tmu}
 \pmb{\upmu}_{GI_n}(\mathcal G)  \equiv -
 {\rm Tr} \left\{
   \hat{\gamma}_{GI_n}(\mathcal G)  \,\hat{\bf r}
 \right\}
 \\
 = (1-\mathsf D_I) 
 \mathcal S[I\leftarrow I_0]\left(
 \pmb{\upmu}_{0n}^{(0)} \right)
 \\
  + \sum_t \sum_{e\ni I}^{\in t}
 \mathcal S[e\leftarrow t]\left(
 \pmb{\upmu}_{00,\pi_I[0n_I]}^{(t)}
  \right)
 \;.
\end{multline}
The required
dimer diabatic transition dipole moments 
can be computed from adiabatic quantities similarly to Eq.~\eqref{e:tdm_ad}. %1-TDMs in AO representation
However, caution is required here. Due to the electronic gauge problem (see Sec.~\ref{sec:gauge}),
adiabatic transition dipole moments obtained from routine electronic structure
programs may not be consistent with the phase of the RM's transition densities.
Therefore, the gauge (here: sign of the 1-TDMs) must be standardized first
and the corresponding transition dipole moments computed from them directly 
by carrying out the trace operation in the AO space,
\begin{equation}
 \pmb{\upmu}_{0\square} = -\sum_{\xi\eta} T_{0\square;\eta\xi} {\bf D}_{\xi\eta} \;,
\end{equation}
where ${\bf D}_{\xi\eta} \equiv  \mel{\chi_\xi}{\hat{\bf r}}{\chi_\eta}$
are the dipole moment one-electron integrals and $T_{0\square;\xi\eta}\equiv 
\mel{\chi_\eta}{\hat{\gamma}_{0\square}}{\chi_\xi}$ is the AO representation
of the 1-TDM.

The theory presented above is especially useful to calculate oscillator strengths
of adiabatic transitions in the aggregate which read as follows 
\begin{equation}
f_a(\mathcal G) = \frac{2}{3}\omega_{a}(\mathcal G) \left|  
  \pmb{\upmu}_{0a}(\mathcal G)
 \right|^2 
 \;,
\end{equation}
where $\omega_{a}$ and $\pmb{\upmu}_{0a}$ 
is the adiabatic transition energy and transition dipole moment,
respectively. The latter is expressed in terms of Frenkel transition dipole moments 
from Eq.~\eqref{e:tmu} as
\begin{equation}
 \pmb{\upmu}_{0a}(\mathcal G) = \sum_{I}\sum_{n\in I} U_{I_na}(\mathcal G) \pmb{\upmu}_{GI_n}(\mathcal G) \;,
\end{equation}
where ${\bf U}(\mathcal G)$ is the aggregate-level diabatic-to-adiabatic 
transformation.
The next sections discuss how to obtain these quantities from
bottom-up construction of the electronic-vibrational Hamiltonian operator
and its diagonalization.

% ----------------------------------------------------------------------
\subsection{Hamiltonian Operator}
\label{subsec:electronic_hamiltonian}
% ----------------------------------------------------------------------

We consider the total Hamiltonian operator given by
\begin{equation}
 \mathcal{H}(\mathcal G) = 
 \hat{T}(\mathcal G) 
 + \mathcal{H}_{\rm el}(\mathcal G)
 \;.
\end{equation}
We first derive approximate expressions for the kinetic energy operator,
and subsequently discuss the electronic Hamiltonian.

% ----------------------------------------------------------------------
\subsubsection{Kinetic Energy}
\label{subsubsec:kinetic_energy}
% ----------------------------------------------------------------------

The kinetic energy operator is formulated by
quantizing the canonical momenta obtained in Eq.~\eqref{e:multiscale_canonical_momenta}.
Here, we choose an approximate linear form
of the kinetic energy that is fairly standard and sufficient
in most chemistry contexts, $\hat{T}(\mathcal G) \cong \frac{1}{2}
{\bf g}^{-1} : {\bf p} \otimes {\bf p}$.
Substituting the canonical momenta with their quantum mechanical
representations it leads to
\begin{multline} \label{e:t}
 \hat{T}(\mathcal G) \cong
 \frac{1}{2}
 \hat{1}_{\rm el}
 \sum_I
  {\bf g}_I^{-1}
   : \hat{\bf p}_I \otimes \hat{\bf p}_I
 %\\
 +
 \frac{1}{2}
 {\bf g}_{\bm{\Xi}}^{-1} :
 \left( \hat{\bf P} - \hat{\bf Y}
 \right)
 \otimes
 \left( \hat{\bf P} - \hat{\bf Y}
 \right)
 + \ldots
 \\
 \approx
 \hat{1}_{\rm el} \hat{T}_0(\mathcal G) 
 %+ \frac{1}{2}\sum_j \left( \sum_I  {\bf K}_{jI} : \left\{ {\bf q}_I \otimes {\bf p}_I + {\bf p}_I \otimes {\bf q}_I \right\}  \right) P_j
 \\
 + \frac{1}{2}\sum_j \sum_I\left(   {\bf K}_{jI} :  \hat{\bf p}_I \otimes \hat{\bf q}_I
    + 
                                    {\bf K}_{jI}^\mathsf T :  \hat{\bf q}_I \otimes \hat{\bf p}_I 
   \right) \hat{P}_j
\end{multline}
where
\begin{equation} \label{e:t0}
 \hat{T}_0(\mathcal G) \equiv
  \frac{1}{2}
 \sum_I
  {\bf g}_I^{-1}
   : \hat{\bf p}_I \otimes \hat{\bf p}_I
 +
 \frac{1}{2}
 {\bf g}_{\bm{\Xi}}^{-1} :
  \hat{\bf P} \otimes
  \hat{\bf P}  \;,
\end{equation}
and ${\bf K}_{jI} \equiv -\sum_k \left\{ {\bf g}_{\bm{\Xi}}^{-1} \right\}_{jk} {\bf A}_{kI}$
are the local Coriolis-like couplings,
which are analogous in spirit to global Coriolis couplings originating
from the removal of global rotation of a mechanical system\cite{Littlejohn.Reinsch.RevModPhys.1997}
(see Appendix~\ref{sec:gauge_potential_collective}).
It is clear that the multiscale design induces
dynamical coupling between the local and collective modes.
Note that quantization of classical phase space variables introduced symmetrization
in Eq.~\eqref{e:t} to preserve hermiticity of the kinetic operator.

% ----------------------------------------------------------------------
\subsubsection{Electronic Hamiltonian}
\label{subsubsec:electronic_hamiltonian}
% ----------------------------------------------------------------------

The MBE-2 expansion of Eq.~\eqref{e:mbe2}
applied to the electronic Hamiltonian
results in
\begin{equation}
 \mathscr{H}_{\rm el}(\mathcal G) = \sum_I (1-\mathsf D_I) \mathscr{H}_{{\rm el},I} 
 + \sum_e \mathscr{H}_{{\rm el},e} 
\end{equation}
and, in a more familiar notation, one can express it also as
\begin{multline}\label{e:ham_general}
  \mathscr{H}_{\rm el}(\mathcal G)
  =
  V_{G}(\mathcal G) \ket{G}\bra{G}  
  \\
  +
  \sum_I \sum_{n\in I} \epsilon_{I_n} (\mathcal G) \ket{I_n}\bra{I_n}
  +
  \sum_I \sum_{nm\in I}^{n\neq m}
  \lambda_{I_{nm}} (\mathcal G) \ket{I_n}\bra{I_m}
  \\
  +
  \sum_{I\neq J} \sum_{n\in I} \sum_{m\in J}
  J_{I_nJ_m} (\mathcal G) \ket{I_n}\bra{J_m}.
\end{multline}
By comparing the two expressions at a fixed nuclear geometry,
the matrix elements valid up to the MBE-2 order can be found to be
\begin{subequations}
 \begin{align}
  \epsilon_{I_n}  (\mathcal G) &= (1-\mathsf D_I)\epsilon_{I_n} + \sum_J (1-\mathsf D_J) v_J + \sum_{e\not\ni I} v_e 
  \nonumber\\&\qquad+
  \sum_{e\ni I} \eta_{e;\pi_I[0n],\pi_I[0n]}
  \;,\\
  \lambda_{I_{nm}}(\mathcal G) &= (1-\mathsf D_I)\lambda_{I_{nm}} 
  \nonumber\\&\qquad+
    \frac{1}{2} \sum_{e\ni I} \left\{ 
   \eta_{e;\pi_I[0n],\pi_I[0m]}
  +\eta_{e;\pi_I[0m],\pi_I[0n]}
  \right\} \;,\\
  J_{I_nJ_m}      (\mathcal G) &= \mathsf A_{IJ} J_{I_nJ_m} \;,
 \end{align}
\end{subequations}
where for $e=(I,J)$ we define
\begin{subequations}
 \begin{align}
   \epsilon_{I_n}       &\equiv \mel{n_I}   {\mathscr{H}_I}{n_I} -v_I   \;,\\
   \lambda_{I_{nm}}     &\equiv \mel{n_I}   {\mathscr{H}_I}{m_I}        \;,\\
   v_e                  &\equiv \mel{0_I0_J}{\mathscr{H}_{IJ}}{0_I0_J}  \;,\\
   \eta_{e;n0,m0}       &\equiv \mel{n_I0_J}{\mathscr{H}_{IJ}}{m_I0_J}  \;,\\
 %\beta_{e;n0,0m}       &\equiv \mel{n_I0_J}{\mathscr{H}_{IJ}}{0_Im_J}  \;.
  J_{I_nJ_m}            &\equiv \mel{n_I0_J}{\mathscr{H}_{IJ}}{0_Im_J}  \;.
 \end{align}
\end{subequations}
with $v_I \equiv \mel{0_I}{\mathscr{H}_I}{0_I}$,
and $\mathsf A_{IJ}$ is the adjacency matrix defined by $\mathsf A_{IJ}=1$ if $(I,J)\in \mathcal E$, 
otherwise $\mathsf A_{IJ}=0$.
Note also that due to the scalar nature of the Hamiltonian operator,
the expression in Eq.~\eqref{e:ham_general} is completely free 
from superimpositions. 

At a fixed geometry, the above form of the Hamiltonian
has the same MBE2-based structure as the
REM method.\cite{Ma.Ma.JPCA.2013}
However, TrMBE formulation embeds the framework in a more general graph-covariant
formalism %introduces the scale-bridging component
making its matrix elements well-defined functions
of the fragmentation graph and its structural realization
upon local and global deformations.
In particular, the microscopic quantities have their counterparts in the macroscopic quantities 
defined by the fragmentation graph hierarchy. For example, the excited-state diagonal energy 
of an isolated monomer, $\epsilon_{I_n}$, defined in a local energy reference relative to $v_I$, 
corresponds to the local excitation energy in the aggregate, 
$\epsilon_{I_n}(\mathcal G)-V_G(\mathcal G)$, with corrections arising from the intermolecular 
interactions and connectivity encoded by $\mathcal G$.
The $V_G(\mathcal G)$ offset naturally arises 
from the absolute energy reference adopted for the aggregate Hamiltonian, 
in contrast to the locally referenced excitation energies of the isolated monomers.

Thus, the complex electronic structure of the aggregate
is connected to its local and global deformations on equal footing.
Another specific feature of the TrMBE is its intrinsic gauge structure which is discussed next.

% ======================================================================
\section{Intrinsic Gauge Problem}
\label{sec:gauge}
% ======================================================================

Even in the frustrationless regime discussed in Sec.~\ref{subsubsec:collective_mode_definition}, 
there exist a representation flexibility
at each level of the multi-scale hierarchy, a property which 
is a direct consequence of Eq.~\eqref{e:hierarchy},
leading to a representation gauge freedom.

% ----------------------------------------------------------------------
\subsection{Base Graph Directionality}
\label{subsec:graph_directionality}
% ----------------------------------------------------------------------

The gauge associated with the lowest scale is 
the graph directionality choice discussed before
and is fairly straighforward to define based on the structure of RDs
as exemplified in Fig.~\ref{fig:aggregate_topologies}(c).
However, it has no effect on the final Hamiltonian and collective mode
definition. To see this, %hanging the directionality gauge induces changes
consider a switch of directionality 
of edge $e=(I\rightarrow J)$ as $\mathcal Pe=(I\leftarrow J)\equiv(J\rightarrow I)$ 
that
induces
\begin{subequations}
\begin{align}
 \mathcal P_e {\bf d}_e &= 
 \begin{pmatrix}
  0 & {\bf 1}^{(J)} \\ {\bf 1}^{(I)} & 0
 \end{pmatrix}
 \cdot 
 \begin{pmatrix}
  {\bf d}_e^{(I)} \\ 
  {\bf d}_e^{(J)}
 \end{pmatrix} 
 =
 \begin{pmatrix}
  {\bf d}_e^{(J)} \\ 
  {\bf d}_e^{(I)}
 \end{pmatrix} 
 \;,\\
 \mathcal P_e \pi_I        &= \pi_J \;,\\
 \mathcal P_e \hat{O}_{IJ} &= \hat{O}_{JI} \;.
\end{align}
\end{subequations}
By applying the above transformations to the 
graph-covariant derivative, the Hamiltonian or any other aggregate-level operator
one can show that they are invariant under the graph directionality as long as
it is swapped consistently and does not generate frustrated directions.

% ----------------------------------------------------------------------
\subsection{Local Directionality}
\label{subsubsec:local_directionality}
% ----------------------------------------------------------------------

While the fragmentation graph gauge is straightforward, 
gauge choices that originate at the local and Hilbert space
levels %, which arise due to lifts to these spaces, 
are less obvious within our model and require more caution.
In order to construct the Hamiltonian from many independent monomer and
dimer electronic structure calculations, one must first define a common phase
convention for both the local vibrational modes and electronic excitations. Since
the calculations of the building blocks are performed independently
of each other, the resulting vibrational frequency eigenvectors 
and electronic eigenstates %TDDFT 
are determined only up to an overall sign/phase.
Therefore
one cannot directly use the calculation results
done on monomers and dimers to construct 1-TDMs from Eq.~\eqref{e:tdm}
or Hamiltonian from Eq.~\eqref{e:ham_general}.

% ----------------------------------------------------------------------
\subsubsection{Local Mode Phases}
\label{subsubsec:local_mode_phases}
% ----------------------------------------------------------------------

Vibrational analysis of monomers in each RD %average dimer
can easily lead to a mismach in eigenvector phases.
This induces a gauge freedom of the eigenvectors,
\begin{equation}
 {\bf L}_I
 \longrightarrow
 {\bf L}_I'
 =
 s_I {\bf L}_I'
\end{equation}
for $s_I = \pm 1$.
Using
vibrational modes with random $s_I$ and superimposing them on dimers would inevitably
lead to an erroneous curvature along the local modes in the aggregate.

% ----------------------------------------------------------------------
\subsubsection{Local Electronic State Phases}
\label{subsubsec:electronic_gauge}
% ----------------------------------------------------------------------

A similar situation arises for electronic states
since for the isolated monomer calculation
there is an ambiguity, for example for the Frenkel basis
we have $\ket{I_n} \rightarrow f_{I_n} \ket{I_n}$.
This induces a gauge transformation of the Hamiltonian
\begin{equation}
 \mathscr{H}_{\rm el}
 \longrightarrow
 \mathscr{H}_{\rm el}'
 =
 \hat{G}\mathscr{H}_{\rm el}\hat{G}
\end{equation}
for 
$\hat{G}:=\mathrm{diag}(f_1,f_2,\ldots,)$.
The diagonal energies are unchanged and the electronic couplings transform as
\begin{equation}
 J_{I_nJ_m}
 \longrightarrow
 J'_{I_nJ_m}
 =
 f_{I_n} 
 f_{J_m} 
 J_{I_nJ_m} 
\end{equation}
and so on.
This sign plays no physical role in a
single calculation, but it becomes important when e.g. transition densities from
different monomers and dimers are compared, projected, or assembled into an
aggregate Hamiltonian.

However, the situation with local electronic states is
more complex than for the local modes.
Even if the local phases of the particular monomers
have been fixed, there still remains an additional sign freedom that
is
induced by the fragmentation graph $\mathcal G$. 
It can be seen by analyzing cycles within $\mathcal G$
and their effect on the sign topology of possible electronic state definitions.
The fragmentation graph
has $n_{\rm cyc} = n_{\mathcal E} - n_{\mathcal V} +1$ independent cycles where
where $n_{\mathcal E}$ and $n_{\mathcal V}$ is the number of edges and vertices, respectively.
For a closed cycle $C$ along the graph edges, we define its sign
as $\phi_C = \prod_{e\in C} {\rm sign}(J_e)$.
This sign is invariant under
the local gauge transformation since
$\phi'_C =
   \prod_{e\in C}
   {\rm sign}(f_I f_J J_{IJ}) \equiv \phi_C $.
The last equality follows from the fact that
each vertex on a closed path occurs twice in this product, hence all factors
$f_I^2$ cancel.

For a chain or spanning tree shown in Fig.~\ref{fig:aggregate_topologies}(a) 
we have $n_{\rm cyc}=0$ which 
means that there is no ambiguity in a sign choice, making
all random sign choices valid and physically equivalent  
as long as the Hamiltonian and other operators are rephased accordingly.
However, for a graph containing cycles such as interacting
chains or networks, there will be $2^{n_{\rm cyc}} > 0$
amount of non-equivalent gauge choices.

To fix this problem,
below we define a global gauge based on
a reference monomer which is superimposed
onto each building blocks to re-generate the reference states
used subsequently in projection-based diabatizations.
This ensures that the correct and consistent electronic gauge
is employed.
Additional considerations appear if there is more than one monomer type, 
but they can be accomodated analogously to the above procedure.

% ----------------------------------------------------------------------
\subsection{Local Gauge Standardization}
\label{subsec:gauge_standard}
% ----------------------------------------------------------------------

Let ${\bf T}_{M_0}$ be the RM's 1-TDM in AO representation computed for some
electronic transition, and its superimposed form on some monomer $J$ is
${\bf T}_{J\leftarrow M_0} = \mathcal S[J\leftarrow M_0] \left({\bf T}_{M_0}\right)$.

\subsubsection{Global Phase for Monomer States}
The phase of the target monomer transition density can be fixed 
by computing the
overlap between ${\bf T}_{J}$ 
and ${\bf T}_{J\leftarrow M_0}$ 
given by the Frobenius product under the AO overlap as a metric,
\begin{equation}
 S_J =
 \left\langle
     {\bf T}_J, {\bf T}_{J\leftarrow M_0}
 \right\rangle_{\bf s} ,
\end{equation}
where
\begin{equation}
 \left\langle
     {\bf A},{\bf B}
 \right\rangle_{\bf s}
 =
 {\rm Tr}
 \left[
     {\bf A}^\dagger \cdot
     {\bf s}_{AB}  \cdot
     {\bf B} \cdot
     {\bf s}_{BA}
 \right] \;,
\end{equation}
and $s_{\xi\eta} \equiv \braket{\chi_\xi}{\chi_\eta}$ is the AO overlap matrix.
The phase factor assigned to monomer
$J$ is therefore
\begin{equation}
 f_J =
 {\rm sign}
 \left( {\rm Re}\,S_J \right)\;.
\end{equation}
The gauge-adjusted monomer transition density is therefore 
${\bf T}_J\rightarrow f_J {\bf T}_J$.
This operation fixes the sign of each monomer excitation relative to the same
reference monomer transition density transported onto the corresponding target
geometry.

\subsubsection{Global Phase for Dimer States}

For an $IJ$-th RD, 
the phase %reference dimer transition density used in the projection procedure 
is generally still arbitrary with respect to the full aggregate, even though the
projection inherently preserves the phase information within a dimer
with a chosen RD state. 
To fix this arbitrariness when assembling the aggregate, the easiest and most natural strategy is to 
assume the RD's diabat is composed of the superimposed RMs, 
say ${\bf T}^{(IJ)}_{I\leftarrow M_0}$
and ${\bf T}^{(IJ)}_{J\leftarrow M_0}$.
The corresponding monomer overlaps are
\begin{subequations}
 \begin{align}
  S_I^{(IJ)} &= \langle
  {\bf T}_I^{(IJ)}, {\bf T}^{(IJ)}_{I\leftarrow M_0} 
 \rangle_{\bf s} \;,\\
  S_J^{(IJ)} &= \langle
  {\bf T}_J^{(IJ)}, {\bf T}^{(IJ)}_{J\leftarrow M_0} 
 \rangle_{\bf s} 
 \end{align}
\end{subequations}
where ${\bf T}_I^{(IJ)}$ and ${\bf T}_J^{(IJ)}$
are initial diabatic 1-TDMs.
Analogously as in the monomer standardization,
the phase-adjusted monomer components of the dimer reference 
are rephased by $f_I^{(IJ)}$ and $f_J^{(IJ)}$.

These two monomer contributions are then 
used to construct the 
gauge-standardized 1-TDM of the RD's diabatic reference
composed of non-interacting monomers.
In the complete neglect of basis set superimposition error (BSSE),\cite{Boys.Bernardi.MolPhys.1970}
i.e., the so called monomer-centered basis set (MCBS),\cite{Chalasinski.Gutowski.ChemRev.1988}
it is just a 
block-local sum of the two consistently phased monomer 1-TDMs in
the dimer AO representation,
\begin{equation}
 {\bf T}_{IJ}^{\text{MCBS}} =
 \begin{pmatrix}
     f_I^{(IJ)} {\bf T}_I^{(IJ)} & {\bf 0} \\
     {\bf 0}    & f_J^{(IJ)} {\bf T}_J^{(IJ)}
 \end{pmatrix},
\end{equation}
when the AO basis is ordered by fragments. 
In the dimer-centered basis set (DCBS)\cite{Chalasinski.Gutowski.ChemRev.1988}
which utilizes the ghost atoms on the other monomer,
\begin{equation}
 {\bf T}_{IJ}^{\text{DCBS}} = f_I^{(IJ)} {\bf T}_I^{(IJ)} + f_J^{(IJ)} {\bf T}_J^{(IJ)} \;.
\end{equation}
The resulting reference states can be now safely used in the 
projection-based diabatizations to obtain electronic and vibronic couplings.

 \begin{figure}
   \centering
   \includegraphics[width=0.45\textwidth]{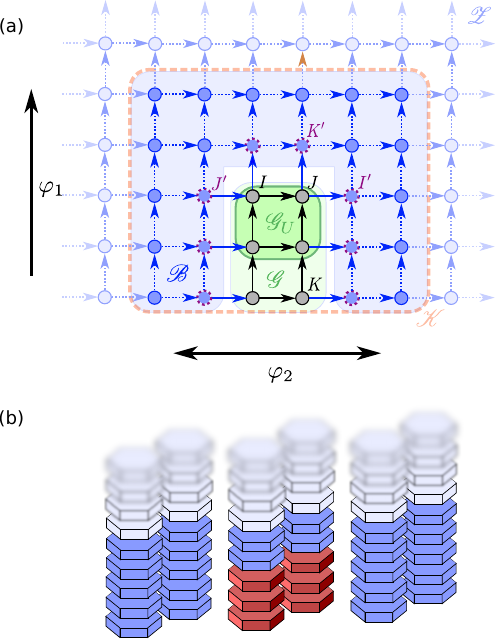}
   \caption{
   \textbf{Illustration of Real Space Approximation to Periodic Boundary Conditions in Extended Systems.}
   The example describes two semi-finite vertical chains assembled into bundles.
   (a) The isometry $\varphi_1$ is applied to the unit cell $\mathcal G_U \in \mathcal G$
   replicating it in one vertical direction.
   Another isometry $\varphi_2$ replicates $\mathcal G_U$ in both horizontal directions.
   The boundary graph space $\mathcal B$ is selected based on the number of single replications
   by each isometry and can be extended. The nearest neighbour boundary $\mathcal B'$
   is relevant for electronic operators leading to closure on a graph hyper-sphere.
   Delocalized spatially-resolved operators such as collective mode vibrations
   or associated vibronic couplings
   require taking into account the full $\mathcal B$ and projecting them from 
   the union $\mathcal K = \mathcal B \cup \mathcal G$ onto $\mathcal G$.
   (b) Example of replicated system lifted to 
   physical space that correspond to the fragmentation graph from (a).
   }
   \label{fig:pbc}
 \end{figure}

\begin{figure*}
   \centering
   \includegraphics[width=0.98\textwidth]{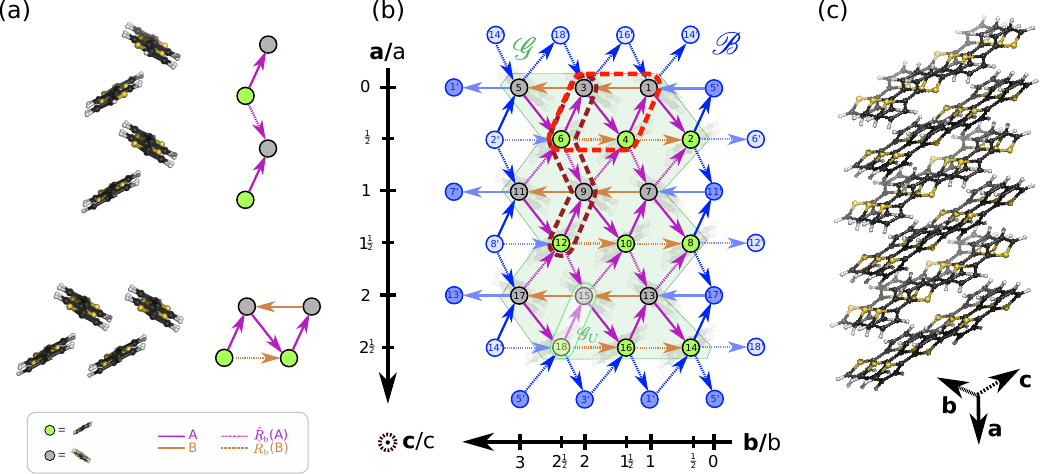}
   \caption{
   \textbf{Graph Representations of Test Systems Based on $\alpha$4T/HT Cutouts.}
   (a) Small cutouts for validation against full QM: 
       \too (above) and \oto (below);
   (b) Larger \TTo cutout as part of crystallographic lattice
       including the two-dimensional PBC setup with minimum redundancy boundary. %$\mathcal B$.
       Crystallographic lattice vectors are shown, with $c$ direction
       ponting back, parallel to the view direction.
       See Fig.~\ref{fig:pbc} for the meaning of graph sub-spaces;
   (c) Side view of \TTo cutout revealing the \textbf{c} axis.
   }
   \label{fig:val_sys}
\end{figure*}

% ======================================================================
\section{Periodic Boundary Conditions}
\label{sec:pbc}
% ======================================================================

In the previous sections we focused on a finite aggregate.
Here we develop the protocol to approximately describe infinite-dimensional
extended aggregates in terms of the self-replicating unit cell
via Periodic Boundary Conditions (PBC) in real space. 
First we formally define the group properties of the PBC
in the fragmentation graph space. Subsequently, we discuss the graph embedding
in the 3D Euclidean space that defines the real (atomistic) structure
of the aggregate. 
Finally, we derive the graph-covariant derivative
operator ${\nabla}^{\mathcal G}$ 
from Sec.~\ref{subsubsec:collective_mode_definition}
that is compatible with the PBCs but is defined in the
real graph space $\mathcal G$. The last result is crucial for the calculation
of the collective vibrational normal modes that are compatible with the PBCs.

\subsection{Group Action Generating PBCs}
Consider an embedding of $\mathcal G$
in physical space such as for example the 3D Euclidean space $E$,
which is schematically shown in Fig.~\ref{fig:pbc}.
We define a free group action $\Gamma$ that generates images
of $\mathcal G$ in physical space
as %$P$.
an isometry in $E$, i.e.,
$\Gamma \subset E(3) $
and we additionally require that it
has no fixed points.
Thus $\Gamma$ is of pure translation type in the graph space, 
as well as can be a pure translation, a screw motion, or other similar isometry
that has no fixed points. 
The entire (infinite and periodic) graph is thus generated via
\begin{equation}
 \mathcal Z = \bigcup_{\varphi \in \Gamma} \varphi(\mathcal G) \;,
\end{equation}
where $\varphi$ is a single extension of $\mathcal G$
that reproduces a part of extended system.
The situation is schematically depicted in Fig.~\ref{fig:pbc}.

\subsection{Minimum Redundancy Boundary}

Consider now a sub-graph 
$\mathcal K$ such that $\mathcal K\subset \mathcal Z$ and 
\begin{equation}
 \mathcal K = \mathcal G \cup \mathcal B
\end{equation}
with $\mathcal B$ being the
boundary space of $\mathcal G$. 
The boundary space contains the minimum amount of
edges (connections) that are generated by $\Gamma$ such that
the full structure of $\mathcal Z$ can be deduced and reconstructed.
In practice, $\mathcal B$ contains connections with at least one virtual node
as shown in Fig.~\ref{fig:pbc}(a).
Application of $\varphi$ to each cell at graph space
generates an extension of the system in physical space
in a chosen direction %with its embedding
depicted in Fig.~\ref{fig:pbc}(b).

% ----------------------------------------------------------------------
\subsection{Hamiltonian under PBC}
\label{subsec:hamiltonian_pbc}
% ----------------------------------------------------------------------

The PBC Hamiltonian is assembled by assigning to each boundary edge the same
dimer contribution as to the corresponding internal edge type. The coupling
between a real monomer $I$ and a periodic image of monomer $J'$ 
(with the prime sign signalling the virtual node) is written as
\begin{equation}
 J_{I,J'}(\mathcal G) =
 J_{\tau(I,J)}(\mathcal G) \;.
\end{equation}
The resulting finite Hamiltonian contains the physical monomers explicitly,
while periodic images enter only through their boundary contributions.
The degree and adjacency matrices are also modified, which includes the connections also
to virtual nodes, taking into account the PBC for the $\epsilon(\mathcal G)$
and $\lambda(\mathcal G)$ contributions.

% ----------------------------------------------------------------------
\subsection{Collective Modes under PBC}
\label{subsec:collective_mode_pbc}
% ----------------------------------------------------------------------

The aggregate's properties such as differential operators or functionals that are 
compatible with the above described PBCs 
can be computed in the extended graph $\mathcal K$ first, 
and subsequently projected onto the $\mathcal G$ sub-graph
with all the redundancies from $\mathcal B$ projected out, 
\begin{equation}\label{e:pbc_proj}
 \Pi_\mathcal G : \mathcal K \rightarrow \mathcal G \;,
\end{equation}
where $\Pi_\mathcal G$ is an orthogonal projection operation and
$ a_\mathcal G = \Pi_\mathcal G(a_\mathcal K) $ is a projection 
of a property $a$.
In particular, the relevant transformation is 
for ${\nabla}^{\mathcal K}$ which is representable by the transformation matrix
${\bf F}(\mathcal K)$ that is computed for the extended graph including the boundary.
Then, ${\nabla}^{\mathcal G} = \Pi_\mathcal G \left( {\nabla}^{\mathcal K} \right)$
is equivalent to computing
\begin{equation}
 {\bf F}(\mathcal G) = {\bf U}(\mathcal G, \mathcal G) \cdot \bm{\Sigma}(\mathcal G, \mathcal G) \;,
\end{equation}
where 
\begin{equation}
 {\bf F}(\mathcal K)[:n_{\mathcal E},:] = {\bf U}(\mathcal G, \mathcal G) 
  \cdot \bm{\Sigma}(\mathcal G, \mathcal G)  \cdot {\bf V}^{\mathsf T}(\mathcal G, \mathcal K)
\end{equation}
is the singular value decomposition (SVD) of the slice of the
matrix representation of ${\nabla}^{\mathcal K}$, that takes only
the real graph edges and leaves out the boundary edges.
Note that the projection operation in Eq.~\eqref{e:pbc_proj}
necessarily depends on the extent of the boundary graph
because the replication is done in real space leading
to an approximation of the periodic property on $\mathcal G$.

% ------------------------------------------------------------
\section{Validation: $\alpha$4T/HT Crystal Cut-Outs}
\label{sec:validation}
% ------------------------------------------------------------

The accuracy of the TrMBE method developed in Sec.~\ref{sec:theory}
was assessed 
based on the high-temperature
polymorph of quaterthiophene crystal $\alpha$4T/HT
as this system has been fairly well understood\cite{Lion.Cocchi.etal.PRM.2024,
Siegrist.Kloc.etal.AdvMat.1998,
Kouki.etal.JCP.2000,Tavazzi.etal.CPL.2005,Lang.etal.SynthMet.1999,Laicini.etal.JLum.2005}
while providing non-trivial
fragmentation topology between the quaterthiophene (4T)
monomer fragments and posing significant computational
bottleneck for larger cutouts. The model systems and their fragmentation
graphs are shown in Fig.~\ref{fig:val_sys}.
Here, the validation calculations focus on collective modes and Frenkel excitation
manifold, whereas local vibrational solvatochromism and
dynamical aspects that involve geometry-dependence and inter-scale connections
will be discussed elsewhere.

The crystalline geometry exhibits the monoclinic P$2_1$/a space group
with two molecules per unit cell. 
The periodic structure was obtained from the NOMAD dataset~\cite{Lion.Cocchi.NOMAD.2023}
associated with Ref.~\citenum{Lion.Cocchi.etal.PRM.2024}, 
based on the experimental crystal structure 
reported by Siegrist et al.~\cite{Siegrist.Kloc.etal.AdvMat.1998}
The periodic coordinates were unwrapped into 
intact molecules and replicated to generate finite
\too and \oto cutouts composed of four 4T monomers shown in Fig.~\ref{fig:val_sys}(a),
as well as a larger \TTo cutout composed of eighteen 4T monomers 
shown in Fig.~\ref{fig:val_sys}(b-c),
with the periodic cell in fragmentation graph space
depicted in Fig.~\ref{fig:val_sys}(b)
by bright green square and label $\mathcal G_U$.
The model systems' geometries were further finalized by superimposing
the 4T monomer optimized by using the $\omega$B97XD/6-31G(d)
method\cite{Chai.HeadGordon.PCCP.2008,Hehre.Ditchfield.Pople.JCP.1972,Hariharan.Pople.TCA.1973,Francl.etal.JCP.1982} 
onto experimental cutouts using the Kabsch method.\cite{Kabsch.ActaCrystA.1976} 
During its geometry optimization,
a restriction to preserve the $C_{2h}$ point group was used
to ensure the planarity of the monomers by imposing dihedral angle constraints
as implemented in \textsc{Gaussian16, Revision A.03}.\cite{Gaussian16A.03}
The associated fragmentation graphs utilizing nearest-neighbour
contacts are shown in Fig.~\ref{fig:val_sys}(a-b) 
and consist of one RM type ($M$) and two RD types ($A$ and $B$),
with an additional symmetry operation $\hat{R}_b$
generating reflected RDs through the $ac$ crystallographic plane,
whose normal is parallel to the $b$ direction.
Thus, the sets from Eqs.~\eqref{e:sets_t} and \eqref{e:sets_s} are 
$\mathcal T_1 = \{M\}$,
$\mathcal R_1 = \{\hat{1}\}$,
as well as
$\mathcal T_2 = \{A,B\}$,
$\mathcal R_2 = \{\hat{1},\hat{R}_b\}$.
RVDMs were computed for each RD
using the generalized reduced-dimensionality vibrational analysis method
by Nagy \emph{et al.}\cite{Nagy.Vikar.Lendvay.PCCP.2018}
and Tóbiás \emph{et al.}\cite{Tobias.Arendas.Csaszar.JCTC.2022},
within either a rigid-body approximation\cite{ZepedaHerrera.etal.PCCP_Pentacene}
contributing exactly 6 RVDMs per RD,
or flexible-body model that additionally
includes arbitrary intramonomer modes that can mix with the rigid-body dimer modes. 
The latter scheme will be referred to the \emph{Flex-n} scheme
with the total number of RVDMs per RD equal $6+2n$ with $n$ lowest-frequency
monomer modes included, analogously to the recent scheme by Soprani \emph{et al.}.\cite{Soprani.etal.JCTC_MMD}

The TrMBE method was implemented in an in-house code %\textsc{LatticeTools}
based on the \textsc{Solvshift}\cite{Solvshift} and \textsc{EOPDev}\cite{EOPDev} fragmentation libraries
that handle superimposition via the rigid molecule algorithm (RMA) and various 
vibrational and AO transformations,
as well as 
%the Python interface to 
the \textsc{Psi4, Version 1.10} quantum chemistry program\cite{Smith.etal.JCP_Psi4}
for handling AO basis sets and electron integrals. 
Harmonic vibrational analysis and (TD)DFT calculations of the building blocks needed for TrMBE
assembly of observables,
as well as validation systems, were performed by using \textsc{Gaussian16}.
1-TDMs were visualized by using the \textsc{Pymol} program.\cite{PyMOL}
See Appendix~\ref{sec:rigid_molecule_algorithm} for the RMA
involving the rotation and symmetry operations of 1-TDMs,
as well as Appendix~\ref{a:flex} for the \emph{Flex-n} model derivation.

%1-TDMs were computed
%via \textsc{Gaussian16} and converted to internal \textsc{Psi4} format.

\subsection{Validation Against Supermolecule Calculations}

First, the smaller \too and \oto cutouts were used for direct comparison
against the full (TD)DFT calculations.
Their topology is complementary, 
featuring a quasi 1D chain (\too) and a quasi 2D network (\oto) 
as shown in Fig.~\ref{fig:coll_val}(a).
The latter contains two cycles in $\mathcal G$ space making it particularly interesting
test case with mechanical frustrations discussed in detail
in Sec.~\ref{subsubsec:collective_mode_definition} 
that are neglected in this work
when reconstructing the collective mode manifold.

\subsubsection{Collective Modes}
\label{subsubsec:validate_collective_modes}

To compare the low-frequency vibrational subspaces that constitute 
collective vibrational modes directly with the full QM calculations
we computed three control quantities:
\begin{subequations}
 \begin{align}
   c_N(\mathcal G) &= 
 \sqrt{N^{-1}  \| {\bf u }^{\mathsf T}_N(\mathcal G) \cdot {\bf u }^{\mathrm{ref}}_N  \|^2_F  }
 \;,\\
   c^{\mathrm{ref}}_N(\mathcal G) &=
 \sqrt{N^{-1}  \| {\bf u }^{\mathsf T}_{\rm all}(\mathcal G) \cdot {\bf u }^{\mathrm{ref}}_N  \|^2_F  }
 \;,\\
   c^{\mathrm{EC}}_N(\mathcal G) &=
 \sqrt{N^{-1}  \| {\bf u }^{\mathsf T}_{N}(\mathcal G) \cdot {\bf u }^{\mathrm{ref}}_{\rm all}  \|^2_F  }
 \;.
 \end{align}
\end{subequations}
In the above,
$c_N$ denotes the root-mean-square inner product (RMSIP) between
the first $N$ EC and reference modes,
$c^{\mathrm{ref}}_N$ is the degree of completeness of the first $N$ reference
modes in the total EC vibrational space, whereas
$c^{\mathrm{EC}}_N$ is the degree of completeness of the first $N$ EC modes
in the total reference vibrational space.
The notation $\| {\bf A} \|$ denotes the Frobenius norm of a matrix defined by $\| {\bf A} \|_F^2 \equiv \sum_{ab} |A_{ab}|^2$.
The results are shown in Table~\ref{tab:vib_subspace} for the \emph{Flex-7}
RM model, whereas more detailed analysis is given in the SI.
The RMSIP value increases with the number of modes included
and quickly reaches a quantitative region of 0.9 at 20 modes for the \too\ cutout
and slightly later at 35 modes for the \oto\ cutout.
The $c^{\mathrm{ref}}_N$ are especially high and above $0.9$ for all mode subspaces shown,
indicating that the TrMBE captures almost completely the correct identity
of the motions given in a reference subspace. Finally, $c^{\mathrm{EC}}_N$ are above $0.94$ in all cases
which means that TrMBE approximation does not contaminate the 
resulting modes with wrong directions.

\begin{table}[htbp]
\centering
\caption{
\textbf{Comparison of the Low-Frequency Vibrational Subspaces Obtained with
EC and full-QM calculations for the \too\ and
\oto\ $\alpha$4T/HT Crystal Cutouts.}\textsuperscript{\emph{a}}
}
\label{tab:vib_subspace}
%\begin{tabular}{c ccc ccc}
\begin{tabular*}{\columnwidth}{@{\extracolsep{\fill}}c ccc ccc}
\hline\hline
& \multicolumn{3}{c}{\too}
& \multicolumn{3}{c}{\oto}
\\
\cline{2-4}\cline{5-7}
$N$
& $c_N$
& $c^{\mathrm{ref}}_N$
& $c^{\mathrm{EC}}_N$
& $c_N$
& $c^{\mathrm{ref}}_N$
& $c^{\mathrm{EC}}_N$
\\
\hline

 5 & 0.600 & 0.919 & 0.947
   & 0.556 & 0.910 & 0.964 \\

10 & 0.721 & 0.944 & 0.965
   & 0.777 & 0.941 & 0.978 \\

15 & 0.846 & 0.960 & 0.973
   & 0.807 & 0.942 & 0.983 \\

20 & 0.907 & 0.965 & 0.978
   & 0.844 & 0.944 & 0.985 \\

25 & 0.918 & 0.967 & 0.981
   & 0.863 & 0.936 & 0.986 \\

30 & 0.928 & 0.968 & 0.983
   & 0.880 & 0.936 & 0.987 \\

35 & 0.931 & 0.968 & 0.985
   & 0.904 & 0.936 & 0.988 \\

40 & 0.936 & 0.968 & 0.986
   & 0.916 & 0.937 & 0.987 \\

46 & 0.937 & 0.967 & 0.987
   & 0.935 & 0.935 & 0.988 \\

\hline\hline
\end{tabular*}

\parbox{\columnwidth}{\footnotesize
\textsuperscript{\emph{a}} The \emph{Flex-7} reduced modes were utilized to obtain collective modes
using the EC method from Sec.~\ref{subsubsec:collective_mode_definition}.
}
\end{table}

\begin{figure}
   \centering
   \includegraphics[width=0.49\textwidth]{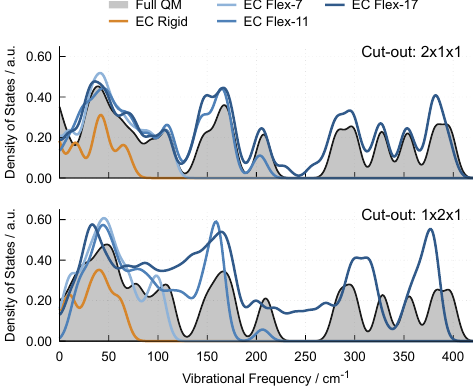}
   \caption{
   \textbf{Raw Vibrational Density of States (VDOS) of Collective Modes using Various Reduced Mode Sets.}
   Null-space threshold was assumed to be 0.25~a.u. in all calculations.
   Reference VDOS is shown in shaded gray plot.
   }
   \label{fig:coll_val}
 \end{figure}

We also studied the performance of various RM models within the TrMBE and its ability to reproduce the
vibrational density of states (VDOS) shown in Fig.~\ref{fig:coll_val}.
Flexible monomer modes must be included to obtain quantitative agreement for
the lowest frequency modes as the rigid body approximation is only qualitatively correct.
Interestingly, adding only the lowest frequency monomer deformations (torsional twists,
in-plane deformations and bending of the chain) substantially improves the VDOS.
Consequently adding more flexible modes between 80--200 cm$^{-1}$ allows to cover higher 
collective mode frequencies. Interestingly, in the case of the \too\ cutout
that is described by a chain graph the VDOS agrees very well with the reference. 
This is not the case for all RM models when \oto\ cutout is considered, whereby
the quantitative agreement is reached only for \emph{Flex-7} model,
whereas higher order RMs lead to visible deterioration of the VDOS that is only qualitative
reproducing the main features but missing the clean frequency windows completely.
We attribute this effect to the frustrationless approximation of the graph-covariant
derivative which is expected to be more sensitive in systems which are described by cycled graphs
such as \oto\ cutout. Nevertheless, the lowest frequency vibrational modes are all quantitatively
described by \emph{Flex-7} model and based on excellent overlap descriptors
from Table~\ref{tab:vib_subspace} in both cutouts we conclude that the collective modes 
are accurately described by the present TrMBE model.

One important comment is regarding the null-space threshold $\lambda^*$ introduced in Sec.~\ref{subsubsec:collective_mode_definition}.
It highly depends on frustrations in the system and other errors,
e.g. caused by superimposition mismatch of RD models onto the system's target structure.
Therefore, in practice a clean drop of eigenvalues was not observed in this study. 
Here, we assumed a constant value of $0.25$ a.u. and this value was not especially optimized.
We also developed an optimization procedure (see the SI for details).

\subsubsection{Excited States}
\label{subsubsec:validate_excited_states}

Now we switch our focus to the excited state manifold in the small cutouts. 
Table~\ref{tab:4t_qm_validation}
shows excitation energies and oscillator strengths of the first four singlet transitions
as compared to reference calculation.
We find the excitation energies in good agreement, with maximal deviation
not exceeding 41~meV. Oscillator strengths are quantitatively reproduced
when dimer species are fully treated as in Eq.~\eqref{e:tmu}.
Interestingly, a common monomer-based approximation, which is equivalent within TrMBE
to remove all edges thus making all $\mathsf D_I=0$,
shows qualitatively a correct picture but considerably overestimates the radiance of the bright states
by 20--35~$\%$. 
We also computed 1-TDMs for all states (see SI) by using Eq.~\eqref{e:tdm}
and we found that the character of the state transitions is quantitatively
reconstructed, with even fine lobe features being reproduced.

\begin{table}[htbp]
\centering
\caption{
\textbf{Comparison Between EC and Full QM TDDFT Calculations for Two
$\alpha$4T/HT Crystal Cutouts Containing Four Monomers.}\textsuperscript{\emph{a}}
}
\label{tab:4t_qm_validation}
%\begin{tabular}{c c cc c ccc}
\begin{tabular*}{\columnwidth}{@{\extracolsep{\fill}}c c c cc c ccc}
\hline\hline
Cutout &
State &&
$E_{\rm EC}$ &
$E_{\rm ref}$ &&
$f_{\rm mon}$ &
$f_{\rm dim}$ &
$f_{\rm ref}$
\\
&
&&
(eV) &
(eV) &&
&
&
\\
\hline

\oto
& 1 && 3.146 & \textbf{3.187}  && 0.0395 & 0.0534 & \textbf{0.0115} \\
& 2 && 3.168 & \textbf{3.194}  && 0.0010 & 0.0001 & \textbf{0.0008} \\
& 3 && 3.305 & \textbf{3.275}  && 0.0007 & 0.0005 & \textbf{0.0008} \\
& 4 && 3.531 & \textbf{3.515}  && 5.0575 & 3.6834 & \textbf{3.7414} \\
\hline

\too
& 1 && 3.139 & \textbf{3.178}  && 0.0016 & 0.0003 & \textbf{0.0005} \\
& 2 && 3.255 & \textbf{3.235}  && 0.1306 & 0.1535 & \textbf{0.0599} \\
& 3 && 3.384 & \textbf{3.346}  && 0.0006 & 0.0006 & \textbf{0.0012} \\
& 4 && 3.472 & \textbf{3.490}  && 4.8768 & 4.0469 & \textbf{4.0091} \\

\hline\hline
\end{tabular*}

\parbox{\columnwidth}{\footnotesize
\textsuperscript{\emph{a}} $f_{\rm mon}$ denotes oscillator strengths
obtained from the monomer-only approximation, in contrast to the full
MBE-2 result $f_{\rm dim}$ from Eq.~\eqref{e:tmu}, where the interactions
between connected dimers are also included.
}
\end{table}

To connect the collective modes and potential energy surfaces together,
we carried out scans along the highest frequency theoretical collective mode in the \oto\ cutout
as a non-trivial example of a system with cycles (see SI, Fig.~S4).
For this purpose, we compared the displaced aggregate adiabatic state PES from supermolecular 
calculations with TrMBE energies computed for $V(\mathcal G_{\Xi_{17}})$ 
as well as $E_a(\mathcal G_{\Xi_{17}})$, where $E_a$ is the $a$th adiabatic state PES
as a function of the displacement of the 17th collective mode, which was calculated
according to Eqs.~\eqref{e:d}-\eqref{e:coll_freq}. Note that once the aggregate
is displaced from ${\bf Q}_0$, its original symmetry is broken and we used unrestrained
summations for each graph edge explicitly. One can see that
both ground and excited state PES are in excellent agreement despite the 2-body
nearest-neighbour approximation being adopted in the chosen FM. Most importantly, the vibrational frequencies
from the full QM and TrMBE scans are 66.2 and 65.5~cm$^{-1}$, respectively,
to be compared with the analytically calculated TrMBE vibrational frequency of that mode 
equal to 65.2~cm$^{-1}$ based solely on the ${\bf Q}_0$ data. Therefore, our analysis indicates that
the analytical curvature around ${\bf Q}_0$ is accurately represented,
thus TrMBE provides means to obtain a reliable collective mode dependence on both ground
and excited states at the same level of theory.

 \begin{figure*}
   \centering
   \includegraphics[width=0.92\textwidth]{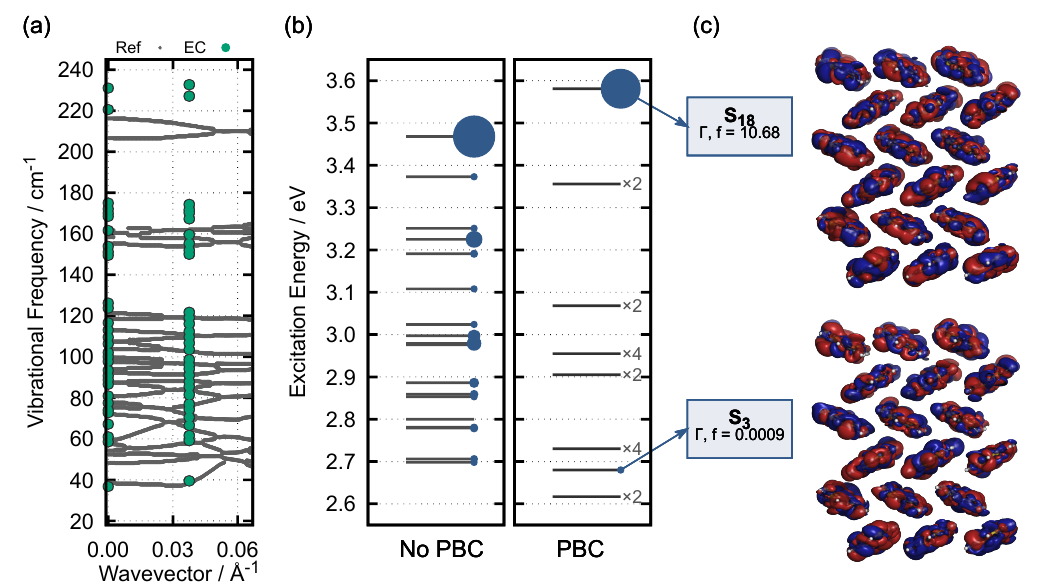}
   \caption{
   \textbf{Collective Modes, Electronic Excitations and One-Particle Transition Densities of \TTo\ Cutout.}
   (a) Vibrational phonons for $\Gamma$ point and $(1/3,0,0)$-point
   computed with PBCs as compared to the band structure for full crystal calculations 
   by Hermet et~al.\cite{Hermet.etal.JPCA_Alpha4TPhonons};
   (b) Electronic excitation energies without and with PBCs.
   The oscillator strength is depicted via blue blobs.
   State degeneracy with PBCs is indicated next to the energy levels;
   (c) 1-particle transition densities for the Dyavidov bright pair
   of non-degenerate $\Gamma$-point transitions.
   The isovalue was set to $\pm~1\times 10^{-4}$ a.u.
   }
   \label{fig:val_331}
 \end{figure*}

\subsection{Optical Exciton Structure of the Periodic 4T Crystal}
\label{subsec:demo}

To further demonstrate the applicability of the TrMBE method,
we used the \TTo cutout
that is a considerable challenge %the computational feasiblity 
for supermolecular 
calculations at the (TD)/$\omega$B97XD/6-31G(d) level of theory.
Fig.~\ref{fig:val_331} shows the periodic phonon vibrations,
optical excitation spectrum as well as bright state 1-TDMs. 

First, we directly compared the selected phonon modes 
at the $\Gamma$-point
as well as at the $(1/3,0,0)$-point 
against reference calculations 
by Hermet et~al.\cite{Hermet.etal.JPCA_Alpha4TPhonons}
who used
the
Perdew, Burke, and Ernzerhof (PBE)
exchange-correlation functional\cite{Perdew.Burke.Ernzerhof.PRL.1996} 
combined with the %The interactions between ions and electrons were
projector augmented wave (PAW) method\cite{Blochl.PRB.1994} for periodic treatment.
We reconstructed these phonon modes 
directly from the TrMBE collective modes computed on a \TTo\ supercell
that were first folded onto the primitive supercell and subsequently localized in the associated reciprocal space
via a procedure very similar in spirit to 
the fragment particle-hole density (FPHD) diabatization method by Wang \emph{et~al.}\cite{Wang.etal.JPCL_FPHD}
(see SI for details).
The results are shown in Fig.~\ref{fig:val_331}(a).
Even though we did not include the periodicity along the $c$-axis,
the ${\bf k}$-localized vibrational bands are in good agreement with reference calculations
validating both the TrMBE collective modes as well as the PBC procedure with minimum redundancy boundary.

Next, we analyzed the Frenkel manifold shown in Fig.~\ref{fig:val_331}(b). TrMBE %correctly 
predicts
the highest Frenkel state to be dominantly bright,
in agreement with optical spectroscopy studies\cite{Lang.etal.SynthMet.1999,Laicini.etal.JLum.2005}
where the
strong concentration of oscillator strength 
in the upper Davydov component is dominant.
However, %with and without PBCs. O
only with PBCs employed
the expected degeneracy of states was reproduced based on crystallographic symmetry subspaces
(see SI for details). Moreover, the non-periodic calculation leads to a distribution of the oscillator
strength over most of the states. In contrast,
the only two non-degenerate states were obtained with PBCs which are bright, with very large
oscillator strength redistribution indicating Davydov splitting.
This is expected since only $\Gamma$-point states are typically bright while other states are dark.
Interestingly, our approximate calculation of the Davydov splitting of 0.9~eV is quite consistent
with experimental values that range between 0.36 to 1.21~eV depending on experimental 
conditions.\cite{Muccini.etal.PRB.2000,Kouki.etal.JCP.2000}
Finally, we computed the 1-TDMs for the bright Davydov pair shown in Fig.~\ref{fig:val_331}(c).
The two densities are delocalized and differ mostly only in phase distribution between monomers
which is expected.

% ------------------------------------------------------------
\section{Conclusions}
\label{sec:conclusions}
% ------------------------------------------------------------

In this work, we developed a hierarchical graph-covariant scale-bridging construction
which
shifts the complexity of the operator description and dynamical
equations of motion (be it fully quantum or quantum-classical)
from the physics of the whole system's identity
to particular choices of 
i) efficient configuration scale division and frames,
ii) efficient electronic structure method
governing particular operators in question at a given scale, and
iii) dynamical level regime at each scale.
In other words, aggregate-level representations
of the operators and dynamical generators acquire
non-trivial geometric connections as well as discrete fragmentation
structure inherited from the lifts of the dual graph representation,
but being in principle computationally tractable via graph assembly across distinct scales.
The general design strategy can be therefore summarized as follows:
 \begin{enumerate} 
   \item {\bf Hierarchy.} Define the scales of the problem and the encapsulation hierarchy $(\bm{\mathcal X}|\bm{\mathcal G})$.
   \item {\bf Frames.}    Choose coordinate frames ${\bf X}$ in each scale (intrinsic gauge).
   \item {\bf Geometry.}  Define constraints at each scale level and recursively establish their geometry, 
                          using the known highest configuration space $\mathcal X^{(1)}$ as a starting point. 
   \item {\bf Operators.} Choose an electronic structure fragmentation method 
                          to construct aggregate-level operators $\hat{O}(\mathcal G)$ 
                          based on calculations on small building blocks that
                          correspond to vertices and edges of $\bm{\mathcal G}$.
   \item {\bf Dynamics.}  Establish equations of motion for the desired dynamical level of theory to propagate the system in time.
 \end{enumerate}
The above general procedure is expected
to preserve %, at least in theory, 
the geometry of the system as much as possible
when deriving the physics of a complex system
across its intrinsic scales.
This may include accounting for global stiffness, elasticity and other properties
that originate at the micro scale but extend over larger scales and are otherwise prohibitive
to attack from the supermolecular approach using quantum chemistry methods.

We used this framework to formulate the transition-density many-body expansion (TrMBE)
of the aggregate's electronic structure
with MBE as exemplary FM and (TD)DFT level of electronic structure theory. 
In TrMBE, the Frenkel
states, electronic Hamiltonian, one-particle density matrices, and transition
dipole moments are assembled consistently from monomer and dimer
transition density matrix elements on the discrete fragmentation graph.
The proposed construction divided degrees of freedom into 
hierarchy of coordinates at three spatial/temporal scales:
 i) low-frequency quasi-acoustic collective vibrational modes with time scales
 of pico-seconds and longer;
 ii) medium- and high-frequency local vibrational modes with ultra-fast sub-picosecond time scales;
 and
 iii) ultrafine electronic motion with sub-femtosecond time scales.
TrMBE utilizes a known effective potential style fragmentation strategy\cite{Ma.Ma.JPCA.2013,Liu.Wang.Ma.CJCP.2021,
Durand.Trinquier.Sanejouand.Biopolymers.1994,Houjou.JCTC_CoarseVibrations,Soprani.etal.JCTC_MMD}
and rigid molecule algorithm\cite{Gordon.Smith.Xu.Slipchenko.AnnuRevPhysChem.2013,Blasiak.Londergan.Webb.Cho.ACR.2017}
but can be applied
to describe Hamiltonians of both ground and excited states of condensed-phase systems
while simultaneously linking local and global deformations of materials and soft matter.
We believe that, due to its operator-level formulation,
TrMBE can be naturally employed in high-dimensional wavepacket simulations
possibly leading to unprecedented treatment of very large systems
at a fully non-adiabatic quantum mechanical level.
Our work in Ref.~\citenum{Blasiak.etal.JCTC_PBI} goes in this direction.
On the other hand, superimposing TrMBE on classical MD trajectories is also
possible, and via definition of canonical hierarchical momenta
TrMBE can also serve at the level of quantum-classical schemes. 
We anticipate that due to the above features, TrMBE generalizes 
a single method's interface across 
a variety of classical, quantum and quantum-classical 
dynamical schemes depending on the problem at hand,
and allows one to study molecular and exciton dynamics in
truly disordered, realistic condensed-phase systems. 
We therefore believe that this contribution is a step towards
bridging the quantum-classical gap problem in condensed-phase simulations of chemical systems.

%\section*{Associated Content}

%\subsection*{Data Availability Statement}
%
%The data underlying this study 
%are available in the article and its Supporting Information. 
%Additional data and the source code used for the 
%calculations are available from the corresponding 
%author upon reasonable request. The code is 
%part of a software framework that has been 
%developed over several years and is currently being prepared for a separate, 
%documented public release elsewhere.

\section*{Supporting Information}

%The Supporting Information is available free of charge at https://pubs.acs.org.
Additional validation against full-QM calculations, including
collective vibrational modes with increasing RVDM dimension, diagonal 1-TDMs
comparisons, scan of ground- and excited-state potential energy
surface along a chosen collective mode in \oto cutout;
analysis of collective mode symmetry under periodic
boundary conditions; determination of optimal null-space threshold $\lambda^*$.

%\section*{Author Information}
%
%\subsection*{Corresponding Author}
%
%Bartosz B{\l}asiak, Institute of Physical and Theoretical Chemistry, Goethe University Frankfurt, Max-von-Laue-Str. 7, 60438 Frankfurt am Main, Germany.
%
%*E-mail: blasiak.bartosz@gmail.com

\section*{Author Contributions}

B.B. conceived and developed the theoretical framework, derived the
formalism, designed and performed the calculations, analyzed the
results, and wrote and revised the original manuscript. I.B. contributed
to the design and conceptualization of this work, specifically with
regard to the role of long-range collective modes and connections to
deformation potential theory, and she critically reviewed and edited the
manuscript. I.B. acquired funding for this research through the Chair of
Theoretical Chemistry at Goethe University Frankfurt, in the framework
of a research program on vibronic coupling effects in exciton transport
in molecular materials. All authors have given approval to the final
version of the manuscript.

\section*{Notes}

The authors declare no competing financial interest.

\section*{Acknowledgments}

This work was funded by Goethe University Frankfurt, whose financial support is gratefully acknowledged.

\appendix

% ......................................................................
\section{Flexible-Body Reduced Modes}
\label{a:flex}
% ......................................................................

Here we describe the rigid and flexible-body construction of reduced modes
that are building blocks for collective mode manifold reconstruction
described in Sec.~\ref{subsubsec:collective_mode_definition}
and utilized in validation calculations in Sec.~\ref{subsubsec:validate_collective_modes}. 
The idea is analogous in spirit to recent work
by Soprani et~al.\cite{Soprani.etal.JCTC_MMD} but the construction follows an earlier
vibrational analysis method in a reduced dimensional space of
arbitrary coordinates by Nagy et~al.\cite{Nagy.Vikar.Lendvay.PCCP.2018} and Tóbiás et~al.\cite{Tobias.Arendas.Csaszar.JCTC.2022}, and we recently applied
its rigid-body approximation to pentacene trimers.\cite{ZepedaHerrera.etal.PCCP_Pentacene}
Therefore, we briefly present it here for completeness.

The construction\cite{Nagy.Vikar.Lendvay.PCCP.2018,Tobias.Arendas.Csaszar.JCTC.2022} 
starts by defining an arbitrary reduced-space coordinate basis ${\bf y}$
and the reduced-space mass matrix
\begin{equation}
 {\bf M}_{\bf y}({\bf r}) = {\bf C}^{\rm T}({\bf y}) \cdot
                            {\bf M}_{\bf r}^{\frac{1}{2}} \cdot
                            {\bf P}^{\rm vib}_{\bf r}({\bf r}) \cdot
                            {\bf M}_{\bf r}^{\frac{1}{2}} \cdot
                            {\bf C}        ({\bf y}) \;,
\end{equation}
where $M_{{\bf r};\alpha u,\beta w}=m_\alpha\delta_{\alpha\beta}\delta_{uw}$,
${\bf P}^{\rm vib}_{\bf r}({\bf r})$ is the vibrational projector in the Cartesian space
and ${\bf C}({\bf y})$ defines the ${\bf y}$-to-Cartesian transformation
\begin{equation} \label{ea:c}
 C_{\alpha u,Ia}({\bf y}) \equiv \left( \frac{\partial r_{\alpha u}}{\partial y_{Ia}} \right)_{\bf y} \;.
\end{equation}
The transformation in Eq.~\eqref{ea:c} is constructed for three translations
and three Euler rotations per monomer (see Ref.~\cite{ZepedaHerrera.etal.PCCP_Pentacene}),
plus optionally $n$ selected (arbitrary) isolated monomer normal modes superimposed
on the monomers in the dimer,
\begin{equation} \label{ea:d}
       \overline{\bf C}_{\alpha u,I}^{\text{\emph{Flex-n}}}({\bf y}) = 
\begin{pmatrix} 
  \delta_{ux} \\ \delta_{uy} \\\delta_{uz} \\
  \left( \frac{\partial r_{\alpha u}}{\partial \vartheta_{1,I}} \right)_{\bf y} \\
  \left( \frac{\partial r_{\alpha u}}{\partial \vartheta_{2,I}} \right)_{\bf y} \\
  \left( \frac{\partial r_{\alpha u}}{\partial \vartheta_{3,I}} \right)_{\bf y} \\
 \left[  \mathcal S[I\leftarrow I_0] \left( {\bf L}^{(I_0)} \right) \right]_{\alpha u, 0} \\
  \vdots \\
 \left[  \mathcal S[I\leftarrow I_0] \left( {\bf L}^{(I_0)} \right) \right]_{\alpha u, n} \\
\end{pmatrix}
\;,
\end{equation}
In Eq.~\eqref{ea:d}, $\vartheta_{a,I}$ are the Euler
angles that define the rotation state of the $I$th fragment for $a=1,2,3$. 
To ensure orthonormality in the reduced space, pre-transformation in Eq.~\eqref{ea:d} is 
subjected to the Gram-Schmidt orthonormalization procedure on its columns
leading to an orthonormal transformation in Eq.~\eqref{ea:c}. Once it is defined,
the full-dimensional Cartesian Hessian is reduced to the ${\bf y}$ subspace, mass-weighted
by ${\bf M}^{-\frac{1}{2}}_{\bf y}({\bf r})$ and diagonalized to obtain
the reduced modes ${\bf d}$ from Eq.~\eqref{e:d} and their vibrational frequencies.
This amounts to $6+2n$ RMs after global translations
and rotations are removed. For more details, refer to Refs.~\cite{Nagy.Vikar.Lendvay.PCCP.2018,Tobias.Arendas.Csaszar.JCTC.2022}

% ......................................................................
\section{Rigid Molecule Algorithm}
\label{sec:rigid_molecule_algorithm}
% ......................................................................

In this work, we base the bottom-up principle of TrMBE method
on the representable dimer approximation within the MBE-2
introduced in Sec.~\ref{subsubsec:average_dimer_concept}.
This approach is %quite 
analogous to fragmentation approaches
developed previously such as EFP2,\cite{Gordon.Smith.Xu.Slipchenko.AnnuRevPhysChem.2013}
and fragmentation methods developed by some of us in particular.\cite{Blasiak.Ritchie.Webb.Cho.PCCP.2016,
Xu.Blasiak.Cho.Layfield.Londergan.JPCL.2018,Blasiak.Maj.Cho.Gora.JCTC.2015,Blasiak.Bartkowiak.Gora.PCCP.2021}
These methods require translating observables or electronic wavefunctions
of reference molecules via the lifts of the superimposition map $\mathcal S$
to the associated ``finer'' space.
For instance, 
EFP2
transfers Fock operators and ground state Slater determinants in LCAO-MO representation, 
as well as
distributed multipole moments
and polarizabilities,\cite{Gordon.Smith.Xu.Slipchenko.AnnuRevPhysChem.2013,Day.Jensen.Gordon.Webb.Stevens.Krauss.Garmer.Basch.Cohen.JCP.1996,Nguyen.Pachter.Day.JCP.2014} 
while
the solvatochromic effective fragment potentials (SolEFP) method
does the same also for the
vibrational derivatives of these quantities.\cite{Blasiak.Londergan.Webb.Cho.ACR.2017,Xu.Blasiak.Cho.Layfield.Londergan.JPCL.2018}
Likewise,
SolCAMM\cite{Blasiak.Lee.Cho.JCP.2013} and TrCAMM\cite{Blasiak.Maj.Cho.Gora.JCTC.2015}
transfer the vibrational electrochromic and excited-state transition
distributed multipole moments, respectively, 
and EOP-TI\cite{Blasiak.Bartkowiak.Gora.PCCP.2021}
effective one-electron operators (EOPs)\cite{Blasiak.Bednarska.Choluj.Gora.Bartkowiak.JCC.2021} %encapsulates
of the Frenkel exciton Hamiltonian at the configuration-interaction-singles (CIS) level.
We shall refer to such an approach as to the ``rigid molecule algorithm'' (RMA).
Specifically here, we employ the RMA to transfer RM's %reference monomer 
and RD's vibrational modes\cite{Blasiak.Lee.Cho.JCP.2013} %, %their transition dipole moments,
and 1-TDMs 
as discussed in more detail in Sec.~\ref{subsec:vibronic_coordinates}. 
The routines were taken from the \textsc{Solvshift} program\cite{Solvshift}
for vibrational eigenvectors superimpositions as well as \textsc{EOPDev} library\cite{EOPDev}
for 1-TDM superimpositions.

\subsection{Superimposition Lift to 1-TDM Space}
\label{subsec:lift_1tdm}

Let $\bm{\varrho}$ and $\bm{\eta}$
be the Cartesian $3\times 3$ unitary
rotation matrix and translation vector generated by the Kabsch superimposition map, i.e.,
 ${\bf Q}_e = \mathcal S[e\leftarrow t]({\bf Q}_t)$
such that ${\bf Q}_e = \bm{\varrho} \cdot {\bf Q}_t + \bm{\eta}$.
The rotation of the 1-TDM in AO space is
generated by the lifted rotation $\bm{\varrho}_{\rm TDM}$ via
${\bf T}_e = \bm{\varrho}_{\rm TDM}^{-1} \cdot {\bf T}_t \cdot \left( \bm{\varrho}_{\rm TDM}^{-1} \right)^{\mathsf T}$.
The elements of $\bm{\varrho}_{\rm TDM}$
are block diagonal, with particular blocks
depending on the AO class, i.e., $s$, $p$, $d$, $f$, etc.
In this work we utilize the Cartesian spherical harmonics expansion up to $f$-type contracted
Gaussian-type orbitals (GTOs), stored in compactified matrix form 
according to the Psi4 AO basis set storage convention.
For the $s$-stype block, the entry is trivial (unity) whereas
for the $p$-stype block the entry is equal to $\bm{\varrho}$.
The non-trivial non-unitary entries for the $d$-type ($6\times 9$)
and $f$-type ($10\times 27$) are given below,
and they
can be used to obtain arbitrary rotated 1-TDM components 
from 1-TDM stored in the compactified matrix form.
For a detailed
derivation refer to the documentation of EOPDev libraries.\cite{EOPDev}

\subsubsection{Rotation of $d$-Blocks}

Define the compactified (6D) and full (9D) multi-index
of the second rank via $\{uw\}$ and $[uw]$, respectively,
where $u,w=x,y,z$.
The associated $d$-block rotation component can be expressed by
\begin{equation}
 \bm{\varrho}^{(d)}_{\{u'w'\},[uw]} = 
 \bm{\varrho}_{u'u}
 \bm{\varrho}_{w'w}
 +
 \Delta_{u'w'}
 \bm{\varrho}_{w'u}
 \bm{\varrho}_{u'w}
\end{equation}
with $\Delta_{uw} \equiv 1 - \delta_{uw}$.

\subsubsection{Rotation of $f$-Blocks}

Define the compactified (10D) and full (27D) multi-index
of the third rank via $\{uwz\}$ and $[uwz]$, respectively,
where $x,y,z=u,w,z$.
The associated $f$-block rotation component can be expressed by
\begin{multline}
 \bm{\varrho}^{(f)}_{\{u'w'z'\},[uwz]} =
 \delta_{w'z'} \left(
 s_{u'w'w'}^{uwz} + \Delta_{u'w'} \left\{ s_{w'u'w'}^{uwz} + s_{w'w'u'}^{uwz} \right\}
 \right)
 \\
 +
 \delta_{u'w'} \Delta_{w'z'}
 \left(
 s_{z'u'u'}^{uwz} +
 s_{u'z'u'}^{uwz} +
 s_{u'u'z'}^{uwz}  
 \right)
 \\ 
 + 
 \Delta_{u'w'} \Delta_{w'z'}
 \big(
 s_{u'w'z'}^{uwz} +
 s_{u'z'w'}^{uwz} +
 s_{w'u'z'}^{uwz} \\+
 s_{w'z'u'}^{uwz} +
 s_{z'u'w'}^{uwz} +
 s_{z'w'u'}^{uwz} 
 \big)
\end{multline}
with $s_{u'w'z'}^{uwz} \equiv \bm{\varrho}_{u'u} \bm{\varrho}_{w'w} \bm{\varrho}_{z'z}$.

\subsection{Inclusion of Symmetry Operations}
\label{a:sym}

Here we describe the generation of 1-TDM that is a result
of application of a reflection operation $\hat{\mathcal R}_{\bf b}$
with its normal along a given direction ${\bf b}$.
First, 1-TDM is rotated via $\hat{\mathcal R}_{\bf b}$ to a local frame
which has a chosen axis (here $y$) parallel to ${\bf b}$.
Next, a diagonal matrix ${\bf U}$ is constructed whose entries
are $U_{\xi\xi}=(-1)^{l_y(\xi)}$
where $l_y(\xi)$ is the angular momentum quantum number of the AO's $y$-component.
Finally, the reflected 1-TDM is transformed back to the original orientation.
The above procedure can be written as
$\hat{\mathcal R}_{\bf b} {\bf T} = \bm{\varrho}_{\bf b} 
\cdot {\bf U} \cdot \bm{\varrho}_{\bf b}^{-1} \cdot 
{\bf T} \cdot (\bm{\varrho}_{\bf b}^{-1})^{\mathsf T}
\cdot {\bf U} \cdot
\bm{\varrho}_{\bf b}^{\mathsf T}$.
If more reflections are needed, the above procedure can be repeated
by modifying the transformation matrix accordingly.

% ......................................................................
\section{Gauge Potential in Collective-To-Local Bundle}
\label{sec:gauge_potential_collective}
% ......................................................................

Here we derive the gauge potential and the associated connection terms ${\bf Y}$
as well as the Coriolis couplings ${\bf K}$
for the fibration structure $\mathcal X^{(2)}\leftarrow \mathcal X^{(1)}$
depicted in Fig.~\ref{fig:geometry}
and expressed in Eqs.~\eqref{e:multiscale_displacement}-\eqref{e:collective_velocity}.
Consider the mapping of scale-bridged displacement to Cartesian space 
from Eq.~\eqref{e:multiscale_displacement},
\begin{equation}
 {\bf Q}(\mathcal G) = {\bf Q}_0 + {\bf u}(\mathcal G) \cdot \bm{\Xi}
    + {\bf L}_{(\bm{\Xi})} \cdot {\bf q}
\end{equation}
and label the global and local charts as $\bm{\Xi}$ and ${\bf q}$, respectively,
with the coordinates assumed to be fully independent of each other (orthogonal).
We also assume the Cartesian metric tensor in $\mathcal X_1$ to have a usual form 
$g_{ab}=\sum_\alpha m_\alpha \partial_a {\bf r}_\alpha \cdot \partial_b {\bf r}_\alpha$.
According to Eq.~\eqref{e:jac-index} and assuming $\nabla_j^{\mathcal G(2)}{\bf u}\approx{\bf 0}$,
the scale-bridging Jacobian is simply
\begin{multline}
 {\bf J}^{(1\leftarrow 2)}_{\bm{\Xi},j}(\mathcal G) \equiv
 \left[ \nabla^{\mathcal G(2)} {\bf Q} (\mathcal G) \right]_j
 \\
 = {\bf u}_j(\mathcal G) + 
 \left[\nabla^{\mathcal G(2)} {\bf L}_{(\bm{\Xi})} \right]_j \cdot {\bf q} \;,
\end{multline}
and the local Jacobian is ${\bf J}_{\bf q}^{(1)}(\mathcal G) \equiv \partial_{\bf q} {\bf Q}(\mathcal G) = {\bf L}_{(\bm{\Xi})}$.
%In the above, 
%%for simplicity %we restrict the final expressions to the reference configuration with ${\bf q}={\bf 0}$,
%%%as well as
%we assumed that
%$\nabla_j^{\mathcal G(2)}{\bf u}\approx{\bf 0}$.
%In the following, for simplicity we further restrict the analysis to 
%${\bf q}={\bf 0}$ as well as $\nabla_j^{\mathcal G(2)}{\bf u}\approx{\bf 0}$.
%%that $\nabla_j^{\mathcal G(2)}\nabla_k^{\mathcal G(2)}
%%{\bf L}_{(\bm{\Xi})} \approx {\bf 0}$ as well as $\nabla_j^{\mathcal G(2)}{\bf u}\approx{\bf 0}$.
%
Therefore, one can define a gauge connection\cite{KobayashiNomizu1963,Littlejohn.Reinsch.RevModPhys.1997} 
as 
\begin{equation}
\hat{\bf A}^{(l)} \equiv \left\{ \nabla^{\mathcal G(l)} {\bf J}(\mathcal G) \right\} \cdot {\bf J}^{+}(\mathcal G) \;,
\end{equation}
where $[\nabla^{\mathcal G(1)}]_s \equiv \partial q_s$ and
the global Jacobian is ${\bf J}(\mathcal G)\equiv \left( {\bf J}^{(1\leftarrow 2)}_{\bm{\Xi}}(\mathcal G), {\bf J}_{\bf q}^{(1)}(\mathcal G) \right)$.
Note that the collumns of ${\bf J}(\mathcal G)$ define the tangent space in the $3N$-dimensional Cartesian configuration space.
Its mass-weighted left pseudo-inverse can be cast 
as
\begin{equation}
 {\bf J}^+ \equiv {\bf g}^{-1}(\mathcal G) \cdot 
 \begin{pmatrix}
   {\bf J}_{\bm{\Xi}}^\mathsf T \cdot {\bf m} \\
   {\bf L}_{\bm{\Xi}}^\mathsf T \cdot {\bf m}
 \end{pmatrix}
\end{equation}
%
%where ${\bf U}_{\bm{\Xi}} \equiv {\bf u} + (\nabla^{\mathcal G(2)} {\bf L}_{\bm{\Xi}}) \cdot {\bf q}$
where ${\bf g}(\mathcal G)\equiv {\bf J}^{\mathsf T}(\mathcal G) \cdot {\bf m} \cdot {\bf J}(\mathcal G)$
and 
\begin{multline}
 {\bf g}^{-1}(\mathcal G) = \\
 \begin{pmatrix}
  {\bf W}   & -{\bf W} \cdot {\bf g}_{\bm{\Xi}{\bf q}} \cdot {\bf g}^{-1}_{{\bf q}{\bf q}} \\
 -{\bf g}^{-1}_{{\bf q}{\bf q}} \cdot {\bf g}_{{\bf q}\bm{\Xi}} \cdot {\bf W} &
  {\bf g}^{-1}_{{\bf q}{\bf q}}  +
  {\bf g}^{-1}_{{\bf q}{\bf q}} \cdot {\bf g}_{{\bf q}\bm{\Xi}} \cdot {\bf W} \cdot {\bf g}_{\bm{\Xi}{\bf q}} \cdot {\bf g}^{-1}_{{\bf q}{\bf q}}
 \end{pmatrix}
\end{multline}
with ${\bf W} \equiv ( {\bf g}_{\bm{\Xi}\bm{\Xi}} - {\bf g}_{\bm{\Xi}{\bf q}} \cdot {\bf g}^{-1}_{{\bf q}{\bf q}} \cdot {\bf g}_{{\bf q}\bm{\Xi}}      )^{-1}$.
Now, by disregarding the off-diagonal blocks of the metric tensor, i.e., 
$ {\bf g}_{\bm{\Xi}{\bf q}} \equiv \left\{{\bf u}^\mathsf T + (\nabla^{\mathcal G(2)} {\bf L}_{\bm{\Xi}})\cdot {\bf q} \right\}^\mathsf T \cdot {\bf m} \cdot  {\bf L}_{\bm{\Xi}}
\approx {\bf u}^\mathsf T \cdot {\bf m} \cdot  {\bf L}_{\bm{\Xi}}   \cong {\bf 0}$ (which is reasonable around the reference point ${\bf q}={\bf 0}$), 
the inverse Jacobian simplifies to
\begin{equation}
 {\bf J}^+ \cong 
 \begin{pmatrix}
 {\bf g}_{\bm{\Xi}\bm{\Xi}}^{-1} \cdot \left\{ {\bf u} + (\nabla^{\mathcal G(2)} {\bf L}_{\bm{\Xi}})\cdot {\bf q}  \right\}^\mathsf T \cdot {\bf m}  \\
 {\bf g}^{-1}_{{\bf q}{\bf q}} \cdot {\bf L}_{\bm{\Xi}}^\mathsf T \cdot {\bf m} \;,
 \end{pmatrix}
\end{equation}
leading to the connections
\begin{subequations}
 \begin{align}
 \hat{\bf A}^{(1)} (\mathcal G) &\cong (\nabla^{\mathcal G(2)} {\bf L}_{\bm{\Xi}})\cdot {\bf g}_{\bm{\Xi}\bm{\Xi}}^{-1} \cdot 
         \left\{ {\bf u} + (\nabla^{\mathcal G(2)} {\bf L}_{\bm{\Xi}})\cdot {\bf q}  \right\}^\mathsf T \cdot {\bf m} \;,\\
     {\bf A}^{(2)} (\mathcal G) &\cong (\nabla^{\mathcal G(2)} {\bf L}_{\bm{\Xi}})\cdot {\bf g}^{-1}_{{\bf q}{\bf q}} \cdot {\bf L}_{\bm{\Xi}}^\mathsf T \cdot {\bf m} \;.
 \end{align}
\end{subequations}
The first connection describes how the collective mode frame changes when 
traveling along a fiber (shown in Fig.~\ref{fig:geometry}(c) by dashed lines),
while the second connection describes how the local mode frame changes
when moving along the base manifold $\mathcal X_2$ (blue surface in Fig.~\ref{fig:geometry}(c)).
Only the second connection is needed for the total time derivative
from Eq.~\eqref{e:multiscale_time_derivative}.
From this, the Coriolis-like coupling matrices
are given by
\begin{equation}
 {\bf K}_{jI}(\mathcal G) = -\sum_k [{\bf g}_{\bm{\Xi}\bm{\Xi}}^{-1}(\mathcal G)]_{jk} \cdot {\bf A}_{kI}(\mathcal G) \;,
\end{equation}
with ${\bf A}_{k}(\mathcal G) \equiv {\bf J}_{\bf q}^+(\mathcal G) \cdot \hat{\bf A}_{k}^{(2)}  (\mathcal G) \cdot {\bf J}_{\bf q}(\mathcal G) 
= {\bf g}_{\bf qq}^{-1} \cdot {\bf L}_{\bm{\Xi}}^\mathsf T \cdot {\bf m} \cdot (\nabla^{\mathcal G(2)} {\bf L}_{\bm{\Xi}})$ .

Finally, we relate here to the lift of the configuration space encapsulation
to Hilbert space described by the orbital picture in Fig.~\ref{fig:geometry}(c).
Analogously to the strictly kinematic situation,
during collective displacements 
local diabatic basis states of each fragment will be dragged along
the path on the base $\mathcal X_2$ inducing an additional connection term
in the Hilbert subspace leading to an effective kinematic connection
\begin{multline}
 \hat{Y}_j^{\mathcal X^{(2)} \leftarrow \mathcal H}(\mathcal G) =
 \hat{Y}_j^{\mathcal X^{(2)} \leftarrow \mathcal X^{(1)}} (\mathcal G)
 \\
 + {\rm i}\hbar \sum_{D_AD_B}
 \ket{D_A}
 \left[ \braket{D_A}{\nabla^{\mathcal G(2)} D_B} \right]_j
 \bra{D_B} \;,
\end{multline}
where $D_A$ and $D_B$ are arbitrary diabatic states of the aggregate
whereas 
$\hat{Y}_j^{\mathcal X^{(2)} \leftarrow \mathcal X^{(1)}}(\mathcal G) = 
\frac{1}{2} \sum_I \big\{ {\bf A}_{jI}(\mathcal G) : {\bf p}_I \otimes {\bf q}_I + {\bf A}_{jI}^\mathsf T(\mathcal G) : {\bf q}_I \otimes {\bf p}_I \big\} $ is
the quantum operator analogue of the classical kinematic connection given
in Eq.~\eqref{e:multiscale_kinematic_connection}.
In practice, the above vector %and scalar %(and to lesser extend scalar)
gauge potentials can be directly used in dynamical scale-bridging schemes
utilizing state-of-the-art FMs,
and the existence and form of these connections emphasizes the role of %general framework
scale-bridging method introduced in this work.

\bibliography{bibliography_r9,EigeneStelle_1_references_r1}

\end{document}